\documentclass[acmtog]{acmart}
\acmSubmissionID{584}
\AtBeginDocument{%
  }
\usepackage{pdfpages}
\usepackage{color}
\usepackage{comment}
\usepackage{soul}
\usepackage{xspace}
\usepackage{fixstyle}
\usepackage{algorithm}
\usepackage{algpseudocode}
\usepackage{multicol, multirow}
\usepackage[subtle]{savetrees}
\usepackage{titletoc}
\newcommand{\methodname}{Garment Particles\xspace}

\definecolor{purple}{rgb}{0.6, 0, 0.6}
\definecolor{darkgreen}{rgb}{0, 0.3, 0}
\definecolor{orange}{rgb}{1, 0.5, 0.}

\newcommand{\ie}{i.e.,}
\newcommand{\eg}{e.g.,}

\expandafter\def\expandafter\normalsize\expandafter{%
    \normalsize%
    \setlength\abovedisplayskip{2pt}%
    \setlength\belowdisplayskip{2pt}%
    \setlength\abovedisplayshortskip{-8pt}%
    \setlength\belowdisplayshortskip{2pt}%
}
\copyrightyear{2026}
\acmYear{2026}
\setcopyright{cc}
\setcctype{by}
\acmConference[SIGGRAPH Conference Papers '26]{Special Interest Group on Computer Graphics and Interactive Techniques Conference Conference Papers}{July 19--23, 2026}{Los Angeles, CA, USA}
\acmBooktitle{Special Interest Group on Computer Graphics and Interactive Techniques Conference Conference Papers (SIGGRAPH Conference Papers '26), July 19--23, 2026, Los Angeles, CA, USA}
\acmDOI{10.1145/3799902.3811102}
\acmISBN{979-8-4007-2554-8/2026/07}
\begin{document}

% \title{Garment Particles: A Unified Garment Flow Model for Unstructured AI-driven Modeling and Design}
%% other options. Feel free to change back.
\title{Garment Particles: A 2D–3D Symmetric Garment Representation for Generation and Editing}
% \title{Garment Particles: A 2D--3D Bidirectional Flow Model for Garment Modeling, Design, and Generation}
% \title{Garment Particles: A Unified Garment Flow Model for Unstructured and Bidirectional Garment Modeling}

\author{Kiyohiro Nakayama}
\orcid{0000-0001-9375-9822}
\affiliation{%
  \institution{Stanford University}
  \country{USA}
}
% \email{w4756677@stanford.edu}
\author{I-Chao Shen}
\orcid{0000-0003-4201-3793}
\affiliation{%
  \institution{The University of Tokyo}
  \country{Japan}
}
% \email{jdilyshen@gmail.com}
\author{Ruofan Liu}
\orcid{0000-0001-9421-7259}
\affiliation{%
  \institution{Institute of Science Tokyo}
  \country{Japan}
}
\author{Yiming Wang}
\orcid{0009-0002-7585-6201}
\affiliation{%
  \institution{ETH Zurich}
  \country{Switzerland}
}
\author{Gordon Wetzstein}
\orcid{0000-0003-4201-3793}
\affiliation{%
  \institution{Stanford University}
  \country{USA}
}

\author{Takeo Igarashi}
\orcid{0000-0002-5495-6441}
\affiliation{%
 \institution{The University of Tokyo}
 \country{Japan}
}

\begin{abstract}
Practical garment design spans two modes: intuitive creation from high-level intent, such as a reference image or text description, and complex low-level editing across 2D sewing patterns and 3D draped geometry, which requires professional training to navigate their complex interdependencies. Yet existing frameworks address only part of this challenge, offering either garment generation from casual inputs or direct editing on sewing patterns.
To support both ends of the spectrum, we propose \methodname, a 5D point-cloud representation that jointly encodes 2D sewing patterns and 3D geometry. This representation enables Garment Particles Flow (GPF), a rectified flow framework that supports intuitive generation from high-level inputs (text, images, sketches) and various editing operations on 2D sewing patterns and 3D geometries via diffusion posterior sampling. Finally, we introduce Particles-to-Pattern Flow that converts generated garment particles into curved-based patterns for simulation. We validate our model's generation ability on multiple datasets, achieving state-of-the-art garment generation results against competitive baselines. Our model also enables many garment editing scenarios, including garment interpolation, sewing pattern editing, point-cloud- and silhouette-conditioned garment generation. Our project website is at \url{https://garment-particles.github.io}.
% We demonstrate a potential application of our model through interactive design interfaces that enable free-form design exploration powered by GPF. 
% Finally, we conduct an extensive ablation study to validate components of our framework.
% To address this issue, we introduce \methodname, a unified garment representation and a corresponding generative model that enables fine-grained bidirectional garment modeling. 
% % driven by a generative prior. 
% Specifically, we represent each garments as a set of 5D point-clouds that jointly encodes its 2D sewing patterns, 3D draped shapes. 
% Then we train a diffusion model that generates these point-clouds either unconditionally or conditioned on text.
% % , which are converted to 2D panel and 3D draped shapes. 
% Using posterior sampling, the trained diffusion model enables various bidirectional garment editing tasks, \eg~2D sewing pattern generation from edited 3D garment and vice versa.
% Additionally, our boundary recovery model reconstructs sewing pattern curve boundaries for downstream tasks.

% More importantly, we demonstrate various garment editing scenarios enabled by our representation. 
% The user study results indicate that our method greatly lowers the barrier to garment editing.
    % The edited point clouds recovered mesh can be edited either in 2D or 3D using existing mesh processing tools. 
    % After the edits, the counterparts are generated via diffusion inpainting. \todo{Our model demonstrates .....}
\end{abstract}



\begin{teaserfigure}
  \includegraphics[width=\textwidth]{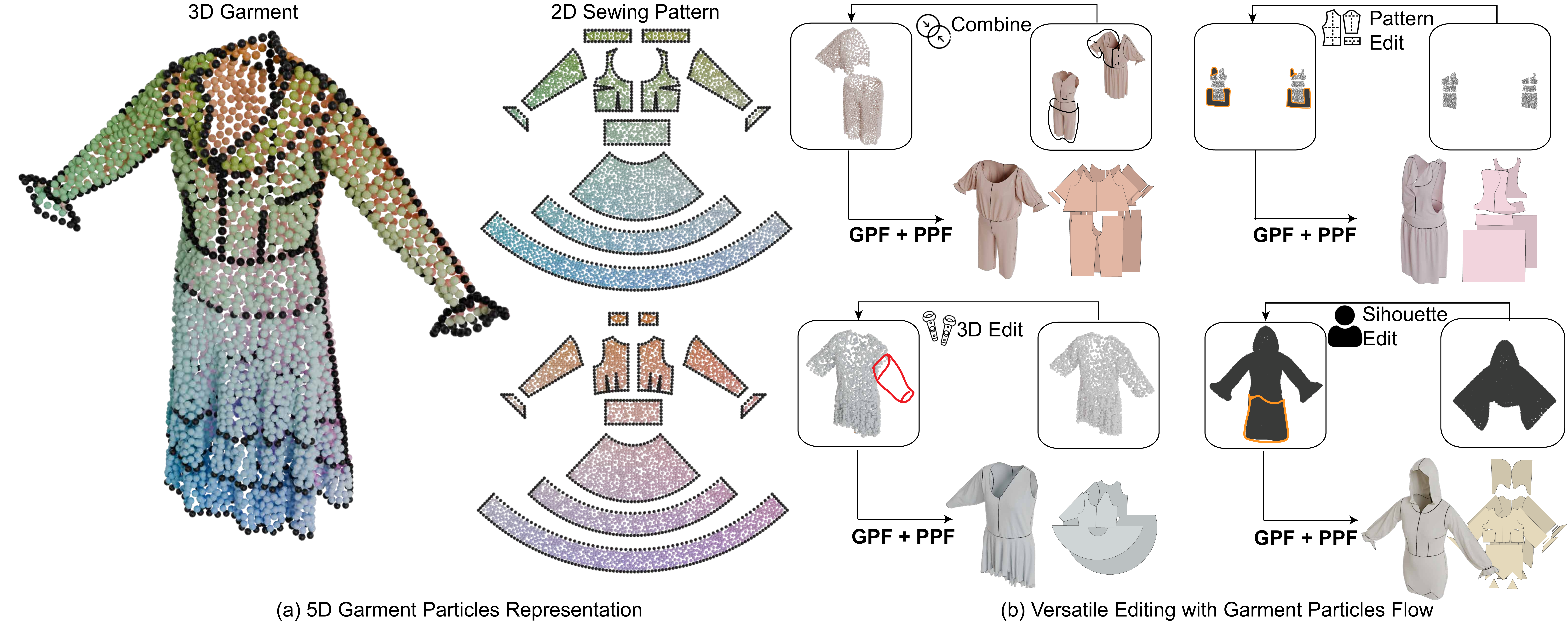}
  \caption{\textbf{Garment Particles} is a garment representation that models both the sewing pattern and its draped garment geometry in a symmetric, 2D-3D point cloud. (a) shows the garment particles representation. The color on the 3D garment (left) and the 2D sewing pattern (right) indicate the same points. \textit{Garment Particles Flow (GPF)}, a generative framework, generates garment particles from multimodal inputs. More importantly, the prior space of GPF enables versatile editing in both 3D garment geometry and 2D sewing pattern domains. Finally, Particles-to-Pattern Flow (PPF) converts the generated particles to simulation-ready sewing patterns. (b) shows the various editing applications enabled by GPF.}
  \Description{\textbf{Garment Particles} is a garment representation that models both the sewing pattern and its draped garment geometry in a symmetric way (left). We present a generative framework, called \textit{Garment Particles Flow (GPF)}, that generates garment particles from multimodal inputs. More importantly, the prior space of GPF enables versatile editing in both 3D garment geometry and 2D sewing pattern domains (right).}
  \label{fig:teaser}
\end{teaserfigure}

\maketitle
\section{Introduction}
Garments are a fundamental aspect of everyday life, yet garment design remains technically demanding and largely left to professionals.  
The primary challenge lies in the pattern-making process, where 2D sewing panels must be carefully shaped to achieve the desired 3D appearance on the body. Professionals often rely on both intuitive creation from reference images or text descriptions and counter-intuitive 2D pattern edits, such as adding darts or adjusting seams, to achieve the desired 3D volume or curvature. 
% Even with design tools such as CLO3D and real-time draping simulations, this non-intuitive, low-level editing remains difficult in pattern making.

In the digital domain, two primary paradigms have emerged to facilitate the garment design process. 
% However, both are incomplete solutions to the rapid and iterative design process in different ways. 
Industry-standard tools, such as CLO3D~\cite{clo3d_website}, Style3D~\cite{style3d_website}, or Marvelous Designer, allow users to directly edit structured, low-level garment representations, including topologically consistent panels and Bézier curves. While precise, these tools require strong pattern-making expertise, as users must understand the complex causal relationship between the geometry of a 2D sewing pattern and its 3D appearance. 
% where a user wishes to experiment with shape without worrying about the underlying mesh structure.
% Separately, data-driven generative models~\cite{nakayama2025aipparel, garmentimage, bian2024chatgarment} enable fast, unidirectional  garment generation from text prompts, sketches, and images, but offer limited support for iterative design. When a user requires fine-grained adjustments, these methods often regenerate the garment, discarding prior results. This could lead to unintentionally altered parts and break consistency with earlier iterations. 

% \ichao{Note: I think one point we can argue is that, we need many different kinds of input for iterative editing. And commonly, it is pretty difficult to combine different control signals. For example, my understanding is that if we encode control signals through LoRAs, combining multiple LoRAs often is not intuitive. We might find more references that support this challenge and perhaps it is better to achieve this through posterior sampling.}
Separately, generative models for sewing patterns~\cite {nakayama2025aipparel, garmentimage, bian2024chatgarment, liu2024multimodallatentdiffusionmodel} create and edit sewing patterns given multimodal, high-level inputs such as texts, images, and 3D scans. While these methods enable rapid pattern development, they rely on modality-specific training, making them ill-suited for garment design, since it often requires different editing tools applied to both the 2D sewing pattern and 3D garment geometry. Further, training a single model to cover all operations is difficult because it would require operation-specific data collection, allocation of training budgets, and careful balancing of the combination of control signals~\cite{he2024dynamiccontrol, wang2025unicombine}.  
% \todo{However, achieving this through combining control signals are typically unintuitive and challenging~\cite{}.} 

To address these limitations, we draw inspiration from the image generation community and cast different garment editing tasks as a training-free inverse problem using diffusion posterior sampling (DPS)~\cite{kim2025flowdpsflowdrivenposteriorsampling, chung2023diffusion, patel2024flowchef}. With DPS, different garment editing applications can be solved by guiding the diffusion sampling process using a specific objective, without retraining the model. However, directly applying DPS to existing garment generative models is challenging, as these models are agnostic to the garment's 3D geometry post-draping. As a result, they learn a 3D-agnostic generative prior that cannot leverage the 3D drape to guide the generated sewing pattern toward a desired draped configuration. The geometry image representation~\cite{li2025garmagenet, geometry_image, yan2024objectworth64x64pixels} addresses this disconnect by rasterizing 2D sewing patterns into images whose opacity encodes the pattern shape and colors the 3D appearance. While this representation captures the 2D--3D duality, it exhibits an \textit{asymmetry} in which the 3D appearance \textit{depends on} the panel shape (\ie~pixel opacity). Consequently, to recover the 3D geometry, a non-differentiable discretization must be performed first to determine the occupied pixels. This makes it difficult to optimize DPS with any objective defined in 3D space directly with respect to the representation itself. 
% As a result, structured tools provide accuracy at the cost of creative flexibility, while generative models offer speed but lack of fine-grained controllability. Neither paradigm adequately support flexible, step-by-step design development.

% \ichao{We can perhaps refer to teaser later because I think we will show representation in teaser.}

In this paper, we propose a novel point-based garment representation, \textit{garment particles}. Garment particles encode both the 2D sewing pattern and its 3D geometry as a 5-dimensional point cloud (\autoref{fig:teaser}a), which can be mapped to the 2D or 3D space via a simple, differentiable projection function. Using garment particles as its underlying representation, we train a flow-based generative model, \textit{garment particles flow} (GPF), and a Particles-to-Pattern Flow (PPF) to generate simulation-ready sewing patterns from multimodal inputs such as sketches, images, and text prompts. 
\begin{figure}[t!]
    \centering
    \includegraphics[width=\linewidth]{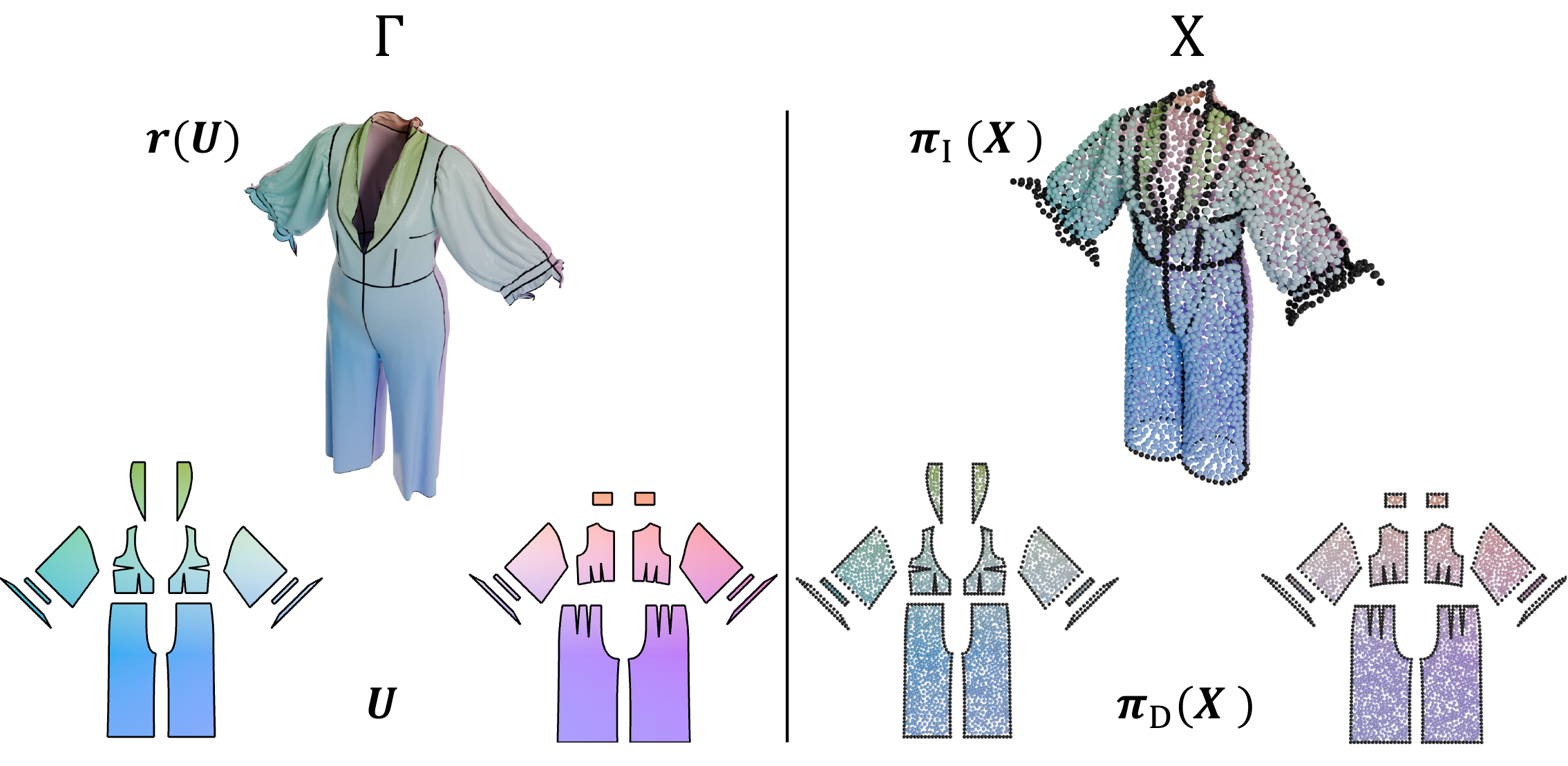}
    \caption{\textbf{Garment Particles Illustration.} \textit{(Left)} We model garments as the graph $\Gamma$ of the parametric function mapping sewing pattern $U$ in $\R^2$ to its draped geometry $\bm{r}(U)$ in $\R^3$. \textit{(Right)} We discretize $\Gamma$ by point samples, denoted as $\bm{X}_{\Gamma}$. Points with the same color in 2D and 3D represent the corresponding points in our representation. Black points mark the boundary of $U$.}
    \label{fig:garment_paticle_illustration}
\end{figure}
% Furthermore, our GPF model learns a semantically rich prior space and is generalizable to different control signals via post-training.  
% compared with baselines. 
More importantly, GPF enables applications of objective-guided sampling techniques, such as DPS, to garments by capturing symmetric relationships between the 2D sewing pattern and its 3D garment geometry.
With GPF, we achieve symmetric, iterative garment editing using a diverse set of tools defined in both 2D pattern and 3D drape spaces. Finally, the generated garment particles can be converted back to simulation-ready sewing patterns using PPF. 

Experimentally, we achieve state-of-the-art generation performance across input modalities and demonstrate the versatility of the GPF prior space across multiple garment-editing tasks. We implemented custom 2D and 3D interactive user interfaces that allow users to casually manipulate 2D sewing patterns, silhouettes, and 3D garment geometries, and use them to guide model generation while maintaining realism and validity.
% Finally, to obtain a simulation-ready curve-based sewing pattern from 5D garment particles, we train a custom particle-to-curve model to generate the sewing pattern curves. We formulate this reconstruction as a conditional generation problem and achieve robust pattern recovery even from noisy garment particles.
% adopt a flow-based diffusion model for this task. 
% \ichao{Do we have any validation to support the following claim?}
% Compared with feedforward reconstruction and training-free approaches, our method achieves the best balance between reconstruction fidelity, particularly for noisy garment particles, which resemble those generated by our GPF model. \todo{To further align the sewing pattern reconstructor to produce a valid pattern, we apply RL post-training with verifiable rewards to further improve the reconstruction quality.} 

Our contributions are:
% \ichao{I think the following content is still too verbose especially if we want to fit into dual-track submission, and have too many repetition with the main introduction text. I have tried to shorten them but I think there is room for improvement.}
\begin{itemize}[nosep]
    \item \textit{Garment Particles.} A novel point-based representation that jointly encodes the 2D sewing pattern and 3D draped geometry of a garment, enabling a fully differentiable pipeline for objective-guided sampling.
    % such as DPS.
    % , enabling a fully differentiable pipeline without boundary discretization. 
    \item \textit{Garment Particles Flow.} A rectified flow model that generates garment particles from multimodal inputs and a particles-to-pattern flow model that recovers simulation-ready sewing patterns from generated garment particles. Our framework achieves state-of-the-art performance in garment generation from texts and images.
    \item\textit{Garment Editing via Diffusion Posterior Sampling.} The learned prior space enables various garment editing applications using DPS, including sewing pattern editing, point-cloud-conditioned garment generation, and silhouette-conditioned garment generation. 
    % supports flexible conditioning on text, images, and sketches. Our approach achieves state-of-the-art performance in both generation and reconstruction, providing a strong foundation for downstream tasks such as guided posterior sampling and latent space interpolation. 
    % \item \textit{Bidirectional Interactive Design Interfaces.} We design a 2D and a 3D user interface that allows users to directly manipulate the garment for GPF-driven garment editing. The set of operations includes painting garment silhouettes from arbitrary views, augmenting or refining sewing patterns, and direct 3D manipulation in a VR environment. 
    % including 2D silhouettes sketching and direct 3D manipulation, enabled by the learned GPF prior.
    % developed a suite of interactive 2D and 3D design interfaces. These tools allow users to define garments by drawing 2D silhouettes or by directly manipulating 3D geometry within a VR environment. These manual edits serve as conditioning signals for the GPF model, which reconstructs complete garment particles satisfying the user's free-formed design intent.
    % \item \textit{RL-Enhanced Pattern Reconstruction.} \ichao{Not so sure whether we want to put this as contribution.} We present a dedicated particle-to-curve model that converts generated garment particles into simulation-ready, curve-based sewing patterns. We employ RL post-training to ensure the generated patterns are topologically accurate and aligned with the input signal.
\end{itemize}

\section{Related Work}
\subsection{Digital Garment Design}
% \ichao{I am thinking about how to put Sensitive Conture in the following structure. Since from my opinion, only our method and Sensitive Conture can achieve this bidirectional editing but using totally different approach. (Please correct me if I am wrong.) Therefore, I think it could be nice to have a separate paragraph discuss this. And I guess the table Takeo mentioned could also be placed and referred to here to support this discussion.}
% Digital garment design has long been studied in graphics research and can be divided into \textit{forward modeling}, which converts sewing patterns into draped garments, and \textit{backward modeling}, which recovers manufacturable sewing patterns from images or 3D scans.
% Forward garment modeling aims to convert sewing patterns into draped garments.
Many automation tools have been developed to accelerate the garment modeling process. Industrial software~\cite{clo3d_website,style3d_website} digitizes the traditional pattern-making process by integrating pattern making and draping simulation. Meanwhile, academic research has focused on reducing manual effort and improving physical fidelity~\cite{auto_seam_allowance, true_seams,GarmentCode2023}. However, these tools require the users to understand the relationship between the sewing pattern and its 3D garment geometry. 

To enable garment design for casual users, research has also focused on automating the pattern-making process. 
% the modeling effort in various aspects of garment design. \cite{auto_seam_allowance} automatically adds seam allowance to sewing patterns so that panels can be physically stitched together. \cite{maneesh_pattern} digitalize sewing patterns from scans of pattern-making books. \cite{true_seams} augments the seam types in digital sewing patterns to allow more realistic garment simulation. 
% To model garments at scale, GarmentCode~\cite{GarmentCode2023} designs a domain-specific language (DSL) for procedurally construct diverse sewing patterns.
% \subsubsection{Backward Garment Modeling}
Early work~\cite{umetani_2011} relies on predefined templates to update 2D patterns from partial 3D edits, whereas later work~\cite {Aric2016, perfect_tailor, Wolff2021DesigningPG, LIU2018113, 10.1145/3306346.3322988, 10.1145/2601097.2601166, 10.1145/2185520.2185532, 10.1016/j.cad.2010.11.008} eliminates the need for templates and enables garment editing via inverse cloth simulation and heuristics. In parallel, prior work also addresses specific components in pattern making, such as dart~\cite{perfect_dart_2023}, pleat placement~\cite{minchen_pleat_2018}, fabric texture alignment~\cite{wolff_2019},  sewing pattern refitting~\cite{chen_refitting_2025, Eggler_2024,MENG201268} and repurposing~\cite{Qi2025rags2riches}. However, these methods are limited in their generalizability to complex garments, leading to unrealistic outputs and requiring long optimization times. These limitations make them unsuitable for practical garment design. 
% Recovering 2D sewing patterns from 3D draped garments is essential for garment editing. 

Another line of work directly recovers sewing patterns from casual inputs. \cite{yang_2018_physics_inspired_garment_from_image, daanen_2008, Sharp:2018:VSC, pietroni2022computational} recovers sewing patterns from images or 3D draped garments using traditional geometric modeling and physics-based simulation.
Recently, deep learning-based methods~\cite{NeuralTailor2022, liu2023sewformer, Li2023isp, li2024garment, dmap_2025, dress123, sewingPCT, wang_2018, garmentimage} enable sewing-pattern recovery from a broader range of input types but are limited to a single modality. To support multi-modal inputs, autoregressive-based~\cite{nakayama2025aipparel, bian2024chatgarment, he2024dresscodeautoregressivelysewinggenerating, zhou2024design2garmentcode, can2026image2garment} and diffusion-based~\cite{liu2024multimodallatentdiffusionmodel, garmentdiffusion, li2025garmagenet} generative models have been adopted. However, these methods rely on a sewing-pattern-centric representation that is decoupled from the final draped garment geometry, making them unsuitable for iterative garment editing that requires symmetric interactions between 2D sewing patterns and 3D garments.

\subsection{Diffusion Models for 3D Generation and Editing}
Diffusion-based 3D generative models have advanced rapidly and employ different representations tailored to downstream applications.
% The most closely related works generate parametric surfaces. 
Specifically, closer to our representation, geometry-image models generate geometry images~\cite{elizarov2024geometry, yan2024objectworth64x64pixels, zhang2025spgen} but ignore surface connectivity. 
BRep-based models~\cite{xu2024brepgenbrepgenerativediffusion, lee2025brepdiff, liu2025hola} rely on low-resolution surface patches and post-processing, losing geometric details and struggling with complex structures on sewing patterns, \eg~darts.
In contrast, \methodname models 2D sewing patterns using a two-stage pipeline. The first stage learns a rich prior space over garment particles that encodes both the 2D sewing patterns and 3D draped geometries as a 5D point cloud. This enables casual generation and symmetric garment editing with DPS. The second stage recovers the sewing pattern from the generated garment particles for downstream applications such as cloth simulation.
\begin{figure}[t]
    \centering
    \includegraphics[width=\linewidth]{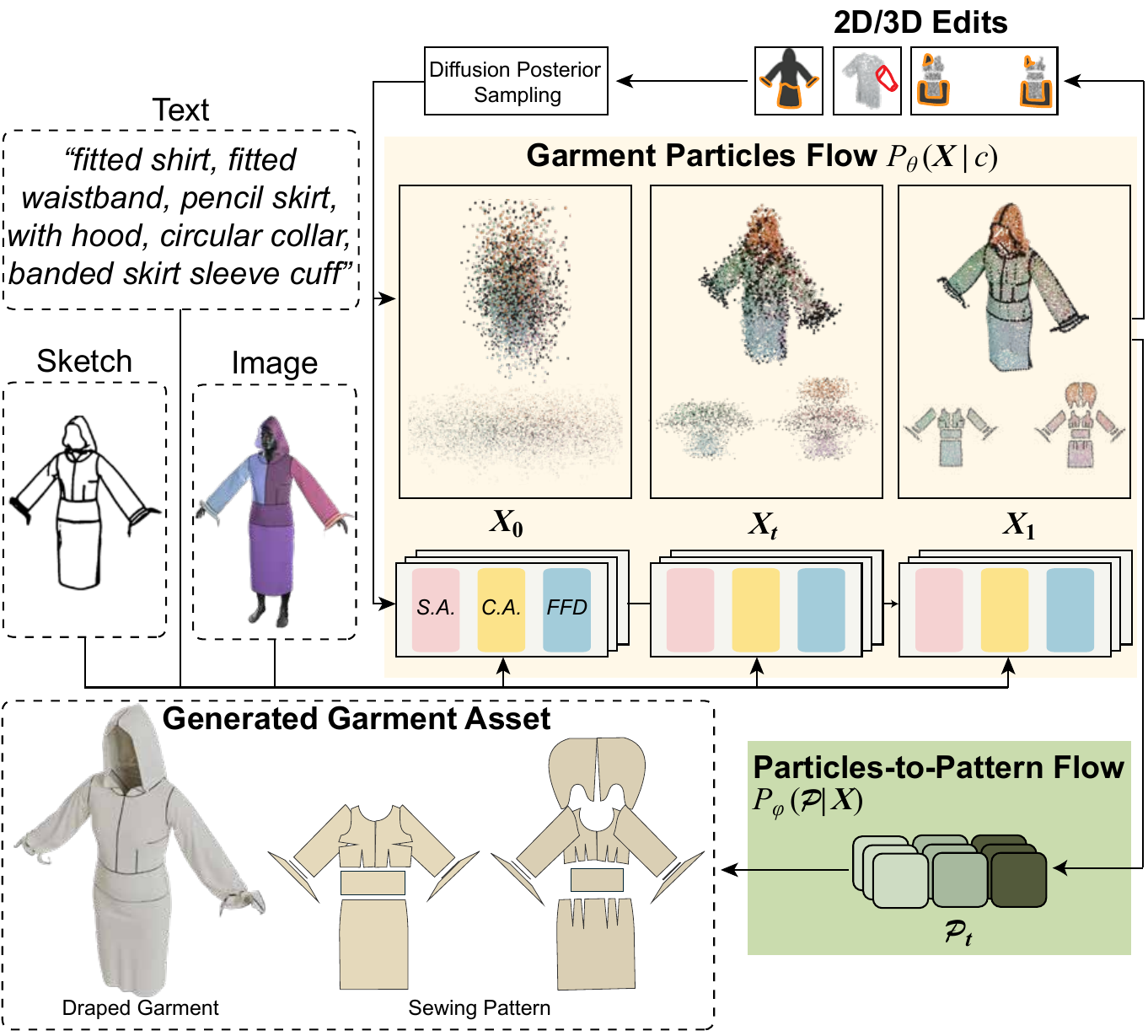}
    \caption{\textbf{Garment Particles Flow (GPF)} is a generative model that generates simulation-ready garments via a two-stage pipeline. In the first stage, multimodal inputs, such as text, sketches, and images, are fed to GPF via cross-attention to generate garment particles $\bm{X}_1$. Diffusion posterior sampling guides the generation based on users' edits. The generated garment particles are then fed into Particles-to-Pattern Flow to generate a vectorized sewing pattern usable for downstream simulation.}
    \label{fig:pipeline}
\end{figure}
Originally introduced for inverse image restoration tasks~\cite{chung2023diffusion}, diffusion posterior sampling (DPS) was used to guide the generation process toward specific objectives. Later works extended this technique to flow models~\cite{kim2025flowdpsflowdrivenposteriorsampling, patel2024flowchef} and to a broader range of generative domains, such as robotics and motion planning~\cite{rempeluo2023tracepace, Yin2025DiverseMP}, medical imaging~\cite{li2024ct}, and audio signal processing~\cite{taufik2025diffusion}. 
We extend FlowDPS~\cite{kim2025flowdpsflowdrivenposteriorsampling} to support diverse garment editing tasks using different guidance objectives.
\section{Garment Particles}
We build a garment representation that encodes both the 2D sewing pattern and the draped 3D garment geometry. Furthermore, the representation should enable both garment-generation and garment-editing applications powered by a learned model. 
To meet these two criteria, we propose \methodname, a garment representation consisting of 5-dimensional particles. Using garment particles, we propose Garment Particles Flow (GPF), a generative framework that can generate and edit garment particles, and convert them to simulation-ready garment assets (\autoref{fig:pipeline}).  

% Furthermore, this representation should enable bidirectional latent space exploration: from ``sewing pattern $\rightarrow$ garment geometry'' and from ``garment geometry $\rightarrow$ sewing pattern''. 

%% ichao: another possible revision.
% To enable interactive garment exploration and editing, we aim to build a unified representation that captures the garment's 3D draped geometry and its 2D sewing pattern. Furthermore, this representation should support latent space exploration, allowing cross-modality editing, i.e., editing 3D draped geometry to update 2D sewing pattern and vice versa. To meet these two design criteria, we represent each garment as a set of 6-dimensional particles, where each can be project both to the spaces of 2D sewing pattern and the 3D garment geometry.  

\subsection{Representing Garment as Particles}
\label{sec:garment_particle_repr}
Mathematically, a cut-and-sew garment can be represented as a 2D parametric surface given by a parametric equation $\bm{r}: U \to \R^3$ where $U \subset \R^2$ is a compact domain. In this setting, the domain $U$ is the sewing pattern, and its image $\bm{r}(U)$ is the 3D garment geometry after draping\footnote{More precisely, $\bm{r}$ also depends on body shape and pose that the garment drapes on. Here we drop these dependencies by assuming the draping is performed on the same body for all garments.}. Therefore, modeling both the sewing pattern and the garment geometry reduces to modeling the graph of $\bm{r}$, which consists of points $\Gamma(\bm{r}) = \set{(\bm{x}, \bm{r}(\bm{x}))}$ for all $\bm{x}\in U$ in the sewing pattern (See left of~\autoref{fig:garment_paticle_illustration}). We can readily recover both the sewing pattern $U$ and its draped 3D garment geometry $\bm{r}(U)$ using the projection operators $\pi_{D}$ and $\pi_{I}$ onto the domain and the image. Such operators are differentiable and computationally simple, making them ideal for garment representation. 

To train a generative model on $\Gamma(\bm{r})$, we discretize it as a 5D point cloud, which we dub the \textit{garment particles} of $\bm{r}$ (See right of~\autoref{fig:garment_paticle_illustration}). Notationally, we use $\bm{X}^{\bm{r}}$ to denote the garment particles of $\bm{r}$ and $\bm{x}^{\bm{r}}$ to denote a point sample from $\bm{X}^{\bm{r}}$. We drop the superscript $\bm{r}$ when there is no ambiguity. 

In practice, we include an additional boundary flag, $f_{\bm{x}}$, indicating whether a point lies on the boundary of a panel to facilitate downstream sewing pattern reconstruction. Concretely, each of our garments $\bm{r}$ is converted to $\bm{X}^{\bm{r}}$ defined as 
\begin{equation}
    \bm{X}^{\bm{r}} = \set{(\bm{x}^{\bm{r}}, f_{\bm{x}^{\bm{r}}}) \,:\, \bm{x}^{\bm{r}}\in \Gamma(\bm{r})}, \quad f_{\bm{x}^{\bm{r}}} = \begin{cases}
        1 & \pi_{D}(\bm{x}^{\bm{r}}) \in \partial U; \\ 
        0 & \text{otherwise.}
    \end{cases}
\end{equation}

\paragraph{Garment Particles Construction}
Given a garment as a mesh, we re-triangulate it with area constraints and use its vertices as point samples. This ensures the point count is roughly proportional to the panel area. To construct $\Gamma$ for each garment, we place its sewing pattern in $\R^2$ without intersection. To ensure semantic consistency across garments, we initialize each panel's location from the 2D projection of its draping initialization transformation, and iteratively resolve panel-wise overlap based on each panel's label (\eg~sleeve, torso, or waistband). 
% See the Supplementary for details about garment particle preparation. 
\autoref{fig:garment_paticle_illustration} shows a packed sewing pattern (\textit{left}) and its associated garment particles (\textit{right}).

\subsection{Garment Particles Flow (GPF)}
We train a rectified flow model to learn a generative prior over garment particles, $P_\theta(\bm{X})$, for garment generation and editing.
% constructed from the GCDv2~\cite{korosteleva2024garmentcodedatadataset3dmadetomeasure}. 
Specifically, we model a probability flow that maps noise $\bm{X}_0 \sim \mathcal{N}\paren{\bm{0}, \bm{I}}, \bm{X}_0\in \R^{N\times 6}$ to garment particles $\bm{X}_1$ from the training data, 
\begin{equation}\label{eq:flow_ode}
    \frac{d\bm{X}_t}{dt} = \bm{v}_{\theta}(\bm{X}_t, t; \bm{c})
\end{equation}
where $\bm{v}_\theta$ models the drift field using weights $\theta$. We can optionally supply $\bm{v}_\theta$ with a conditioning signal $\bm{c}$, which may take the form of text or images. In rectified flow, we assume linear interpolation between any pairs of $\paren{\bm{X}_0, \bm{X}_1}$. 
\begin{equation}
    \bm{X}_t = t\bm{X}_1 + (1 - t)\bm{X}_0 \quad \frac{d\bm{X}_t}{dt} = \bm{X}_1 - \bm{X}_0.
\end{equation}
Thus, the learned drift field $\bm{v}_\theta$ should approximate $\bm{X}_1 - \bm{X}_0$ as much as possible. This motivates us with the standard flow matching loss 
\begin{equation}\label{eq:flow_loss}
    \mathcal{L}_{\text{flow}} = \norm{\bm{X}_1 - \bm{X}_0 - \bm{v}_\theta(\bm{X}_t, t; c)}^2_2
\end{equation}
for training. We use Diffusion Transformers (DiT)~\cite{peebles2023scalablediffusionmodelstransformers, yao2025vavae} as our model architecture. Because the garment particles are unordered, we eliminate positional encodings in the DiT. We set a maximum number of points to 8192 and use masking during training. During inference, the number of points is provided as input, controlling the complexity of the generated sewing pattern.

\paragraph{Injecting Text Condition}
We train GPF to optionally take an input text prompt to guide generation. Specifically, given a text prompt $\mathcal{T}$, we first encode it using CLIP~\cite{radford2021learningtransferablevisualmodels} to obtain text embeddings $\bm{c} \in \R^{77\times 768}$. 
Then $\bm{c}$ is linearly mapped to the same latent space as the GPF model and injected into $\bm{v}_\theta$ via cross attention following~\cite{xiang2024structured}.

% \paragraph{Implementation}
% \ichao{It is ambiguous why we need to separate this with the previous section?}
% We use the garment particle dataset constructed from GCDv2, to train our text-conditioned GPF model. 
We use the LightningDiT-XL variant from~\cite{yao2024fasterdit, yao2025vavae}, which consists of 28 layers of transformer blocks. See the supplementary for additional training details. 

\paragraph{Extending GPF to Image Conditions}\label{sec:image_cond_method}
After text-conditioned training, GPF learns a generalizable prior space that can be easily extended to other modalities without training from scratch. We follow~\cite{ye2023ip-adapter, zhang2024clay} and extend GPF to accept images as an extra condition by adding an extra cross-attention in each transformer block of GPF. The images are tokenized with a frozen DINOv2~\cite{oquab2023dinov2} encoder and then attended to via cross-attention. We initialize the image-conditioned training from the pre-trained, text-conditioned GPF and fine-tune all layers for $160{,}000$ iterations.
% with 8$\times$H100 GPUs.
% We use the same learning rate of 0.0001 but reduce the learning rates of all pre-trained modules by a factor of 10. 

\begin{figure*}[th]
    \centering
    \includegraphics[width=\textwidth]{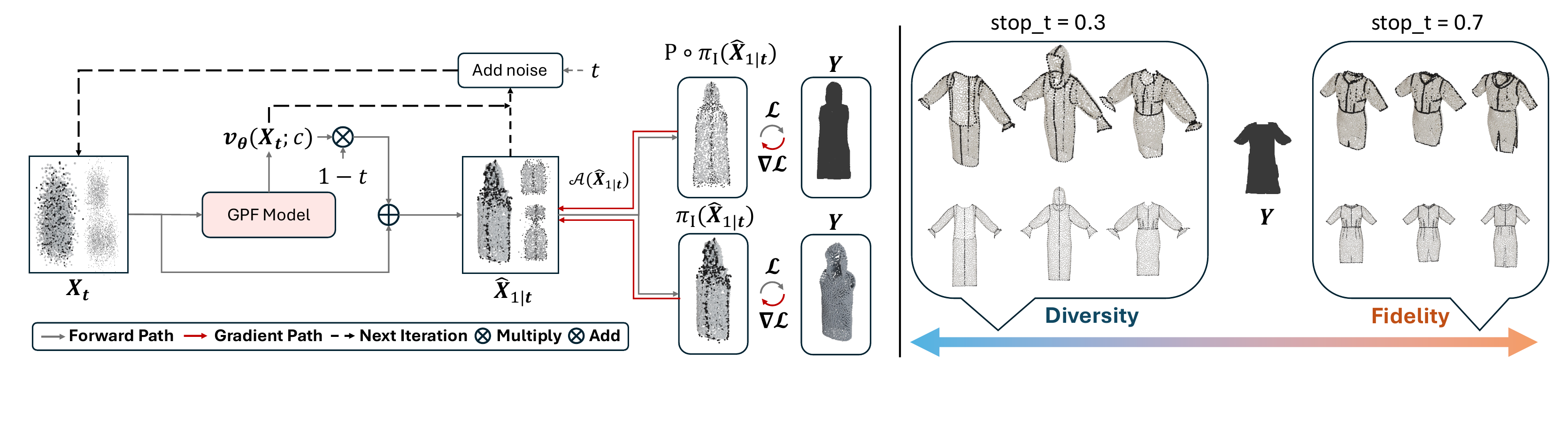}
    \caption{\textbf{Objective Guided Interactions}. \textit{(Left)} By leveraging a trained GPF model, we can optimize the posterior mean $\hat{\bm{X}}_{1|t}$ at each step against an observation $\bm{Y}$ and guide the generation process towards a garment sample that minimizes our specified objective $\mathcal{L}$. \textit{(Right)} By adjusting the hyperparameter $\texttt{stop\_t}$, our objective-guided sampling can produce more faithful (higher $\texttt{stop\_t}$) or more diverse (lower $\texttt{stop\_t}$) results with different random noise input.}
    \label{fig:dps_illustration}
\end{figure*}

\subsection{Recovering Sewing Patterns from Garment Particles} 
It is nontrivial to recover simulation-ready sewing patterns from generated garment particles, as the generation process may be corrupted by noise.
% , due to the noise from the generated samples. 
% To obtain high-fidelity sewing patterns for downstream applications, 
To address this issue, we design Particles-to-Pattern Flow to convert garment particles to a curve-based sewing pattern.
% and outputs a  in the GarmentCode~\cite{GarmentCode2023} format -- including a parametric boundary representation for each panel and stitching relationships defined on panels' edges. 
% \subsection{Vectorized Sewing Pattern Representation}
Specifically, we represent a sewing pattern $\mathcal{P}$ as a tensor of shape $M_{max}\times(E_{max}+1)\times D$ following~\cite{he2024dresscodeautoregressivelysewinggenerating, nakayama2025aipparel, liu2024multimodallatentdiffusionmodel}. 
A pattern consists of a maximum of $M_{max}$ panels, each containing up to $E_{max}$ ordered parametric edges (cubic Bézier curves or arcs). We set $M_{max}=E_{max}=37$ to cover all patterns in our dataset and represent each panel $P_i$ as:
\begin{equation}
    P_i = \operatorname{Stack}\big((T_i, R_i),\; e_1, \dots, e_{E_{\max}}\big)
\end{equation}
where $(T_i, R_i) \in \mathbb{R}^6$ encodes the panel's draping pose (translation and Euler angles), and each edge $e_j \in \mathbb{R}^{15}$ contains information such as control points ($\mathbf{c}_1, \mathbf{c}_2\in \mathbb{R}^{4}$), displacement from previous endpoint ($\delta\mathbf{x} \in\mathbb{R}^2$), arc flag ($f_{\text{arc}}\in\mathbb{R}$), stitching flag ($f_{\text{stitch}}\in\mathbb{R}$), stitch tag ($\boldsymbol{\tau}\in\mathbb{R}^{3}$), boundary condition type ($\mathbf{t}_{\mathrm{attach}}\in\mathbb{R}^{3}$), and validity mask ($f_{\mathrm{valid}}\in\mathbb{R}$). 
% is structured as in the table below. 
% resulting in $D = 15$

% \begin{center}
% \begin{tabular}{lcp{5cm}}
% \toprule
% Component & Dim & Description \\
% \midrule
% $\delta\mathbf{x}$ & 2 & Displacement from previous endpoint \\
% $\mathbf{c}_1, \mathbf{c}_2$ & 4 & Control points (B\'{e}zier or arc) \\
% $f_{\text{arc}}$ & 1 & Arc flag \\
% $f_{\text{stitch}}$ & 1 & Stitching flag \\
% $\boldsymbol{\tau}$ & 3 & Stitch tag (3D landmark for pairing) \\
% $\mathbf{t}_{\text{attach}}$ & 3 & Boundary condition type \\
% $f_{\text{valid}}$ & 1 & Validity mask \\
% \bottomrule
% \end{tabular}
% \end{center}
We formulate sewing pattern recovery as a generation task conditioned on garment particles $\bm{X}$. We model this conditional distribution $P_{\varphi}(\mathcal{P} | \bm{X})$ using another flow model $\bm{v}_{\varphi}$. Unlike GPF, which generates an unordered set of points, sewing patterns consist of ordered edges and panels. Therefore, we use panel and edge embeddings to order the input tokens and employ cross-attention to condition the network. 

PPF learns the garment-particles-to-sewing-pattern mapping purely from data. Although it does not enforce hard constraints—such as boundary points lying exactly on panel edges or interior points staying inside panels—PPF empirically offers the best trade-off between robustness to input noise and reconstruction accuracy\footnote{See Supplementary for a detailed analysis}.

% architecture and training details.

\section{GPF-driven Garment Editing}\label{sec:editing_method}
After training GPF, we can leverage its prior distribution $P_\theta(\bm{X})$ for various \emph{training-free} garment editing tasks. 

\subsection{Garment Interpolation}
A quick way to generate a diverse set of garments is via interpolation in the prior space $P(\bm{X}_0) = \mathcal{N}(\bm{0}, \bm{I})$ and passing the interpolated noise through GPF. We use spherical linear interpolation (SLERP) for noise interpolation. Because the particles are unordered, we follow~\cite{lee2025brepdiff} and first compute a linear assignment between the particles by summing pairwise distances across multiple denoising timesteps. Because GPF requires specifying the particle count per garment, we linearly interpolate the particle count for each intermediate generation when the endpoints' point counts differ. 

% Specifically, Given two noises $\bm{X}_0^{(0)}$ and $\bm{X}_0^{(1)}$ sampled from $P(\bm{X}_0) = \mathcal{N}(\bm{0}, \bm{I})$, we can obtain the interpolated noise $\bm{X}^{(s)}$ where $s\in [0, 1]$ using 
% \begin{equation}
%     \bm{X}_0^{(s)} = \frac{\sin((1-t)\Omega)\bm{X}_0^{(0)} + \sin(t\Omega)\bm{X}_0^{(1)}}{\sin(\Omega)},\, \Omega = \cos(\bm{X}_0^{(0)}\cdot \bm{X}_0^{(1)}). 
% \end{equation}

\subsection{Objective Guided Editing}\label{sec:dps}
% \ichao{I suggest to add a illustrative figure to explain this idea, more specifically, the ``bidirectional'' completion that allows us for achieving different task. Currently, I feel that if some readers do not have the experience in diffusion posterior sampling, it could be difficult for them to imaging how our method works. And I think this kind of figure is vital for graphics paper.}
To enable different garment editing tasks without dedicated training, we use diffusion posterior sampling (DPS)~\cite{kim2025flowdpsflowdrivenposteriorsampling, patel2024flowchef, chung2023diffusion}. DPS solves inverse problems of the form 
\begin{equation}\label{eq:optimal}
    \bm{X}^\star = \operatorname*{arg\,min}_{\bm{X}\sim P_\theta(\bm{X})}\mathcal{L}\paren{\mathcal{A}(\bm{X}), \bm{Y}}.
\end{equation}
Here $\mathcal{L}$ is the objective function, $\mathcal{A}$ is the forward transformation, and $\bm{Y}$ is the observation. Given an observation $\bm{Y}$ which can be derived from some sample $\bm{X}$ via transformation $\mathcal{A}$, DPS seeks the closest sample from our trained prior distribution $P_\theta(\bm{X})$ that minimizes the objective function $\mathcal{L}$ after applying $\mathcal{A}$. Adapted from FlowDPS~\cite{kim2025flowdpsflowdrivenposteriorsampling} for garment particle editing tasks, our sampling algorithm is summarized in~\autoref{alg:dps}.
By adjusting hyperparameters (\texttt{stop\_t}, \texttt{opt\_n}, and $T$), we can balance sample diversity and fidelity (\autoref{fig:dps_illustration}). By varying the conditioning vector $\bm{c}$ across different prompts, we can impose additional control over the generation while minimizing the objective function. 
% To illustrate the versatility of garment editing, 
We demonstrate four garment editing operations in the following. See Supplementary Material for specific hyperparameter settings.
% with their settings shown below. 
% , see the supplementary material. 

\begin{algorithm}
\caption{Diffusion Posterior Sampling of GPF}
\label{alg:dps}
\begin{algorithmic}
\Require Trained GPF $\bm{v}_\theta$, Objective function $\mathcal{L}$, Transformation $\mathcal{A}$, Sampling steps $T$, Observation $\bm{Y}$, Conditioning $\bm{c}$, Guidance Stop Time \texttt{stop\_t}, Learning rate $\eta$, Optimization Steps \texttt{opt\_n}. 
\State $\bm{X} \sim \mathcal{N}(\bm{0}, \bm{I}) \;\; \Delta t \gets \frac{1}{T}$
\For{$\delta  = 0, 1, 2, \dots, T-1$}
    \State $t \gets \frac{\delta}{T}, \;\; \bm{v} \gets \bm{v}_\theta(\bm{X}, t;\bm{c})$
    \State $\hat{\bm{X}}_{0|t} \gets \bm{X} - t \bm{v}, \;\; \hat{\bm{X}}_{1|t} \gets \bm{X} + (1 - t)\bm{v}$
    \If{$t \leq \texttt{stop\_t}$}
        \For{$n = 1, \dots, \texttt{opt\_n}$}
            \State $\hat{\bm{X}}_{1|t} \gets \hat{\bm{X}}_{1|t} - \eta \nabla_{\bm{\hat{\bm{X}}_{1|t}}}\mathcal{L}\paren{\mathcal{A}\paren{\hat{\bm{X}}_{1|t}}, \bm{Y}}$
        \EndFor
        \State $\bm{\varepsilon} \sim \mathcal{N}(\bm{0}, \bm{I})$
        \State $\bm{X}_{0|t} \gets \sqrt{t + \Delta t}\bm{X}_{0|t} + \sqrt{1 - t - \Delta t}\bm{\varepsilon}$
    \EndIf
    \State $\bm{X} \gets (t + \Delta t)\hat{\bm{X}}_{1|t} + \paren{1 - t - \Delta t}\hat{\bm{X}}_{0|t}$
\EndFor
\end{algorithmic}
\end{algorithm}
% \ichao{You mention ``Compared to the original FlowDPS algorithm'', do you mean similar or unlike??}
% \ichao{It is a bit unclear how important the tradeoff you mentioned here. And also, what is ``by chaning the conditioning to different prompt''?}

\paragraph{Point-cloud-conditioned Garment Generation.} Given the 3D garment geometry as a point cloud, we can generate a suitable sewing pattern that matches the input post-draping. We formulate this as a DPS task:
% as follows. 
\begin{equation}\label{eq:dps_recon}
    \begin{split}
        \mathcal{A}&= \pi_{I}: \R^{N\times 6} \to \R^{N\times 3}\\
        \mathcal{L}(\bm{Y}_1, \bm{Y}_2) &= \operatorname{EMD}(\bm{Y}_1, \bm{Y}_2),\; \bm{Y}_i\in \R^{N\times 3}.
    \end{split}
\end{equation}
Here, $\mathcal{A}$ is a projection function that maps the 5D garment particles $\bm{X}_{\bm{r}}$ to the image $\bm{r}(U)$ and $\operatorname{EMD}$ is the Earth Mover Distance~\cite{earthmover}. The generated garment particles will approximate the given 3D geometry when projected to the image domain with $\pi_I$.

% \begin{figure}[t]
% \centering
% \includegraphics[width=0.8\linewidth]{figures/fabrication_example.png}
% \caption{\textbf{Fabrication Examples.} We fabricated generated sewing patterns.
% % generated from our pipeline.
% }
% \label{fig:fabrication}
% \end{figure}
\begin{figure*}[t!]
  \centering
  \includegraphics[width=0.8\textwidth]{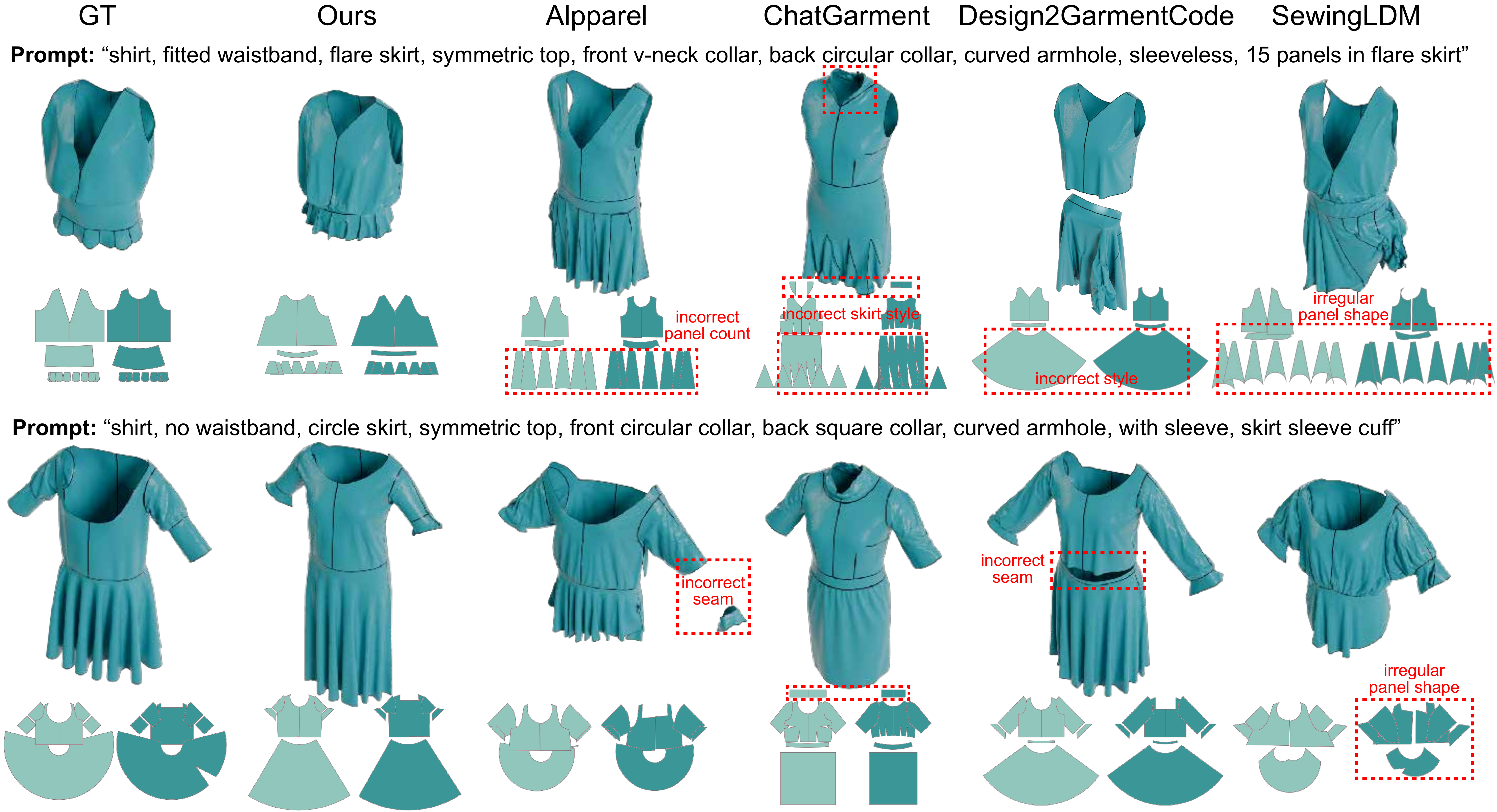}
  \caption{\textbf{Text-conditioned Garment Generation.} The baselines exhibit artifacts, as indicated by the red boxes (e.g., incorrect panel shapes or styles). In contrast, our method outputs realistic garments that align with the input prompt.}
  \label{fig:text_comparison}
\end{figure*}

\paragraph{Garment Completion.}\label{sec:garment_completion}
% \ichao{It is a bit unclear whether the incomplete garment is in 3D or in 2D, or both? If it is in 3D, then the difference from the previous task is that we can complete both 2D and 3D simoutaneously?}
Given an incomplete 3D garment geometry as a point cloud, we can complete it using DPS. For this task, we set 
\begin{equation}\label{eq:one-way-chamfer}
\begin{split}
    \mathcal{A}&= \pi_I:\R^{N\times6} \to \R^{N\times 3}\\
    \mathcal{L}(\bm{Y}_1, \bm{Y}_2) &= \sum_{\tilde{\bm{y}}\in \bm{Y}_2}\min_{\bm{y}\in \bm{Y}_1}\norm{\tilde{\bm{y}} - \bm{y}}^2_2.
\end{split}
\end{equation}
Here, we set the objective function to the one-sided Chamfer Distance, which encourages $\bm{Y}_2$ to lie within $\bm{Y}_1$. In this way, the generated garment particles incorporates the observation $\bm{Y}$ into its geometry, thereby completing the 3D garment geometry in the image while generating its sewing pattern in the domain. 
\paragraph{Sewing Pattern Editing}
Thanks to our symmetric representation, we can also edit the 2D sewing pattern and obtain the resulting garment with our model. By replacing $\mathcal{A}$ in~\autoref{eq:dps_recon} and~\autoref{eq:one-way-chamfer} with $\pi_D$, we can leverage DPS to reconstruct 3D garment geometry from a coarse or incomplete sewing pattern as the observation. Compared with directly editing vectorized sewing patterns, our method automatically generates stitching and draping initialization cues using our particle-to-pattern module. 
\paragraph{Silhouette-conditioned Garment Generation from Arbitrary Views.} Given the silhouette of a garment taken from view $P\in\R^{3\times2}$, we can guide our generation with DPS to sample $\bm{X}$ that has a similar silhouette when projected to $P$. This is done by setting 
\begin{equation}
\begin{split}
    \mathcal{A}&= P\circ\pi_I:\R^{N\times6} \to \R^{N\times 2}\\
    \mathcal{L}(\bm{Y}_1, \bm{Y}_2) &= \operatorname{EMD}(\bm{Y}_1, \bm{Y}_2),\; \bm{Y}_i\in \R^{N\times 2}.
\end{split}
\end{equation}
$\mathcal{A}$ here is the composition of the coordinate projection $\pi_I$ and the view projection $P$, mapping the garment particles into the camera view space. For the objective function, we use EMD in $\R^2$ to encourage the alignment of the generated and observed silhouettes.

\begin{table}[!t]
% \small
    \centering
    \renewcommand{\arraystretch}{0.9}% Tighter
    \caption{\textbf{Text-conditioned Garment Generation.} }
     \resizebox{\linewidth}{!}{%
    \begin{tabular}{l ccc cc cc}
        \toprule
        \textbf{Method} & \textbf{COV}$\uparrow$ & \textbf{MMD ($\times10^3$)} $\downarrow$ & \textbf{1-NNA} $\downarrow$ & \textbf{P-FID} $\downarrow$ & \textbf{CLIP} $\uparrow$ &  \textbf{SSR}$\uparrow$ \\
        \midrule
        Omage& 40.6 & 7.70 & 68.1 & 68.8 & 24.26 & -- \\
        D2GC& 27.6 & 6.29 & 81.2 & 18.9 & \textbf{27.15} & 85.4 \\
        AIpparel & 24.6 & 8.21 & 87.6 & 89.4 & 24.11 & 70.7 \\
        ChatGarment& 13.5 & 8.90 & 88.4 & 16.2 & 25.71 & \textbf{99.5} \\
        SewingLDM & 40.1 & 5.29 & 62.7 & 3.69 & 26.17 & 81.3 \\
        % \highlight{GarmageNet} & \underline{47.9} & \underline{5.09} & \underline{59.9} & \underline{3.47} & 23.51 & -- \\
        \midrule
        \textbf{Ours} & \textbf{48.4} & \textbf{4.64} & \textbf{54.6} & \textbf{3.15} & \underline{26.42} & \underline{91.7} \\
        \bottomrule
    \end{tabular}
    }
    \label{tab:text_gen}
\end{table}

% \begin{table*}[ht]
% % \small
%     \centering
%     % \begin{adjustbox}{width=\textwidth, center}
%     \renewcommand{\arraystretch}{0.9}% Tighter
%     \begin{tabular}{l ccc cc cc}
%         \toprule
%         \textbf{Method} & \textbf{COV} (\%) $\uparrow$ & \textbf{MMD} ($\times 10^3$) $\downarrow$ & \textbf{1-NNA} (\%) $\downarrow$ & \textbf{P-FID} $\downarrow$ & \textbf{P-KID} ($\times 10^3$) $\downarrow$ & \textbf{CLIP} $\uparrow$ &  \textbf{SSR} (\%) $\uparrow$ \\
%         \midrule
%         Omage& 40.6 & 7.70 & 68.1 & 68.8 & 290 & 24.26 & -- \\
%         D2GC& 27.6 & 6.29 & 81.2 & 18.9 & 23.9 & \textbf{27.15} & 85.4 \\
%         AIpparel & 24.6 & 8.21 & 87.6 & 89.4 & 282 & 24.11 & 70.7 \\
%         ChatGarment& 13.5 & 8.90 & 88.4 & 16.2 & 42.9 & 25.71 & \textbf{99.5} \\
%         SewingLDM & \underline{40.1} & \underline{5.29} & \underline{62.7} & \underline{3.69} & \underline{5.22} & 26.17 & 81.3 \\
%         \midrule
%         Ours & \textbf{48.4} & \textbf{4.64} & \textbf{54.6} & \textbf{3.15} & \textbf{3.66} & \underline{26.42} & \underline{91.7} \\
%         \bottomrule
%     \end{tabular}
%     % \end{adjustbox}
%     \caption{\textbf{Text-conditioned Garment Generation.}}
%     \label{tab:text_gen}
% \end{table*}
\begin{figure*}[t!]
\centering
  \includegraphics[width=0.9\linewidth, trim={0 46cm 0 0},clip]{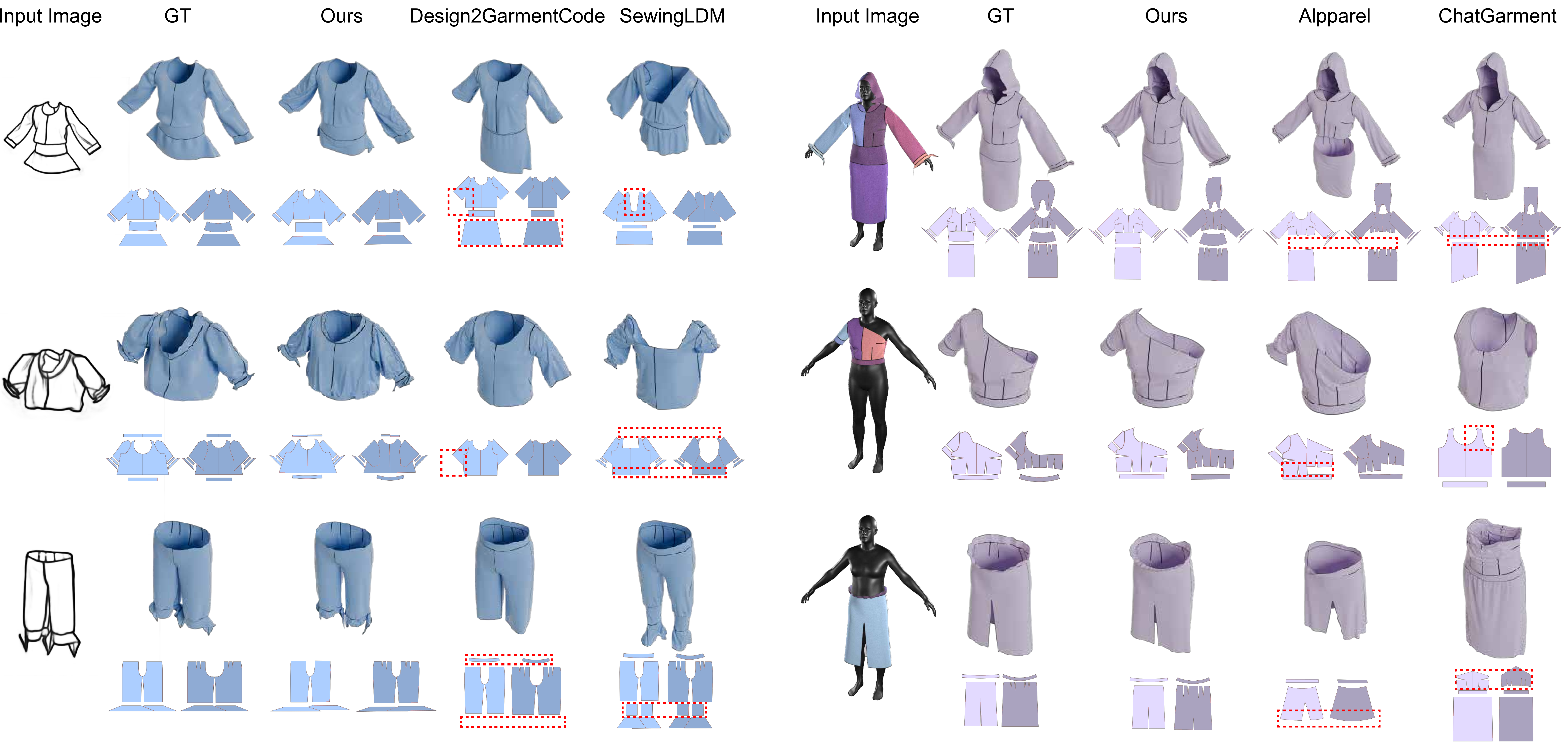}
  \caption{\textbf{Image-conditioned Garment Generation.} Compared to the baselines, which exhibit incorrect pattern style and stitching, our method correctly generates a sewing pattern that yields a draped garment matching the input image for both sketch and GCDv2 image inputs.}
  \label{fig:image_gen}
\end{figure*}
\section{Garment Generation Results}
\paragraph{Dataset}
We evaluate our model's generation performance on GarmentCodeDatav2 (GCDv2)~\cite{korosteleva2024garmentcodedatadataset3dmadetomeasure}.
% , which comprises more than 130k garments with their sewing patterns. 
We obtain the garment particles of each garment in GCDv2 as described in~\autoref{sec:garment_particle_repr}. In total, we obtained $124{,}339$ samples. We randomly split them into training and validation in a 9:1 ratio. To compute all metrics, we randomly sample $1{,}024$ garments from the validation set. 
\paragraph{Baselines}
% Because the final product of our method is sewing pattern, 
We compare our method against various multi-modal sewing pattern generation baselines, including AIpparel~\cite{nakayama2025aipparel},  ChatGarment~\cite{bian2024chatgarment}, SewingLDM~\cite{liu2024multimodallatentdiffusionmodel}, and Design2GarmentCode (D2GC)~\cite{zhou2024design2garmentcode}.
% , all of which are state-of-art methods for sewing pattern generation from multimodal inputs. 
We also compare with geometry-image-based methods Omages~\cite{yan2024objectworth64x64pixels} by adapting for garment generation. 
% as part of our evaluation, to compare with geometry image-based baselines. 

% \begin{figure*}[th]
%     \centering
%     \includegraphics[width=\textwidth]{figures/garment_visual_v1.png}
%     \caption{Text-based Garment Generation}
%     \label{fig:garment_visual}
% \end{figure*}

% After training the GPF model and the \todo{sewing pattern recovery model}, we are able to generate diverse and high quality simulation ready garments either unconditionally or with text conditioning. In this section, we showcase the generated results and evaluate our generation quality against state-of-the-art baselines. 

\subsection{Text-based Generation}
We evaluate the generation diversity and fidelity of our method when using text as control. We constructed the captions procedurally, leveraging GCDv2's design parameters, which specify the garments' design details.\footnote{See Supplementary Material for construction details.}  To ensure a fair comparison, we retrain AIpparel, SewingLDM, and Omages on our captioned dataset. For ChatGarment and Design2GarmentCode, we use their released checkpoints as they use LLMs for text-prompt conversion.

\paragraph{Evaluation Metrics.} 
We measure the 3D garment generation quality with a diverse set of metrics: \textit{3D distribution metrics} (Coverage (COV) score, Minimum Matching Distance (MMD), 1-NN classification Accuracy (1-NNA), and pointcloud-FID (p-FID)), simulation success rate (SSR), and CLIP score~\footnote{See the Supplementary Material for metric calculation details.}.

\paragraph{Results.} 
As shown in~\autoref{tab:text_gen}, our method achieves the best scores on distribution metrics, demonstrating that the generated garment assets are the most diverse and most similar to the test set distribution. This performance boost can be attributed to our 2D--3D representation that captures the symmetric relationship between a sewing pattern and its draped geometry in 3D. By contrast, none of the baselines store 3D information in their representations, making their generation agnostic to the 3D shape after draping. While ChatGarment and Design2GarmentCode slightly outperform our method in simulation success rate and text alignment due to direct GarmentCode generation using LLMs, their outputs are less diverse, reflected by lower coverage score and higher 1-NN accuracy. \autoref{fig:text_comparison} shows qualitative comparisons of the generated garments on different text prompts. The baselines exhibit artifacts, including text misalignment and incorrect seams, panels, and garment styles. In comparison, our model generates high-quality sewing patterns while matching the text descriptions.
% specified in the text prompts.

% \begin{table*}[ht]
%     \centering
%     % \begin{adjustbox}{width=\textwidth, center}
%     \renewcommand{\arraystretch}{0.9}% Tighter
%     \begin{tabular}{l l cc cc cc}
%         \toprule
%         \textbf{Condition} & \textbf{Method} & \textbf{Panel Acc} $\uparrow$& \textbf{Panel IOU} $\uparrow$ & \textbf{Edge Acc} $\uparrow$ & \textbf{Stitch Acc} $\uparrow$ & \textbf{SSR} $\uparrow$ & \textbf{CD} $\downarrow$ \\
%         \midrule
%         \multirow{4}{*}{GCDv2} & AIpparel & 75.49 & 75.47 & 70.36 & 64.67 & 80.57 & 16.3 \\
%         & ChatGarment & 5.86 & 62.67 & 44.43 & 37.72 & \textbf{95.12} & 55.6 \\
%         \cmidrule{2-8}
%         & \textbf{Ours} & \textbf{83.01} & \textbf{77.54} & \textbf{78.20} & \textbf{69.47} & \underline{89.84} & \textbf{7.0} \\
%         \hline
% \multirow{4}{*}{Sketch} & SewingLDM & 20.61 & 54.47 & 50.48 & 43.49 & 80.18 & 54.2 \\
%         & D2GC & 21.48 & 58.92 & 40.41 & 57.18 & \textbf{97.95} & 32.9 \\
%         \cmidrule{2-8}
%         & \textbf{Ours} & \textbf{81.25} & \textbf{76.07} & \textbf{77.05} & \textbf{66.86} & \underline{92.97} & \textbf{8.7} \\
%         \bottomrule
%     \end{tabular}
%     % \end{adjustbox}
%     \caption{
%     \textbf{Image-conditioned Garment Generation.}}
%     \label{tab:image_gen}
% \end{table*}

\begin{table}[!t]
    \centering
    \renewcommand{\arraystretch}{0.9}
    \caption{\textbf{Image-conditioned Garment Generation.}}
    \resizebox{\linewidth}{!}{%
    \begin{tabular}{l c cc cc}
        \toprule
         & \textbf{Panel Acc} $\uparrow$ & \textbf{Panel IOU} $\uparrow$ & \textbf{Stitch Acc} $\uparrow$ & \textbf{SSR} $\uparrow$ & \textbf{CD} $\downarrow$ \\
        \midrule
        \multicolumn{6}{l}{\textit{GCDv2}} \\
        \midrule
        AIpparel & 75.49 & 75.47 & 64.67 & 80.57 & 16.3 \\
        ChatGarment & 5.86 & 62.67 & 37.72 & \textbf{95.12} & 55.6 \\
        \textbf{Ours} & \textbf{83.01} & \textbf{78.20} & \textbf{69.47} & \underline{89.84} & \textbf{7.0} \\
        \midrule
        \multicolumn{6}{l}{\textit{Garment Sketches}} \\
        \midrule
        SewingLDM & 20.61 & 54.47 & 43.49 & 80.18 & 54.2 \\
        D2GC & 21.48 & 58.92 & 40.41 & \textbf{97.95} & 32.9 \\
        \textbf{Ours} & \textbf{81.25} & \textbf{77.05} & \textbf{66.86} & \underline{92.97} & \textbf{8.7} \\
        \bottomrule
    \end{tabular}
    }

    \label{tab:image_gen}
\end{table}
\subsection{Image-based Generation}
We extend GPF to image-based conditioning using the method described in~\autoref{sec:image_cond_method}. To evaluate across different image styles, we fine-tune GPF on three image datasets. (1) \textit{GCDv2}: we use the front and back garment renderings from GCDv2 following~\cite{nakayama2025aipparel}. (2) \textit{Garment Sketches}: we follow~\cite{liu2024multimodallatentdiffusionmodel} to convert GCDv2 garment rendering to sketches using~\cite{pdc-PAMI2023}. (3) \textit{Realistic Garments}: We use the dataset from~\cite{bian2024chatgarment}, which contains realistically textured garments and humans in different poses. For evaluation, we select in-the-wild garment images from 4DDress~\cite{wang20244ddress} and Fashionpedia~\cite{jia2020fashionpedia}.
    % , which are in-the-wild garment captures.  
For GCDv2 and Garment Sketches, we train a multiview GPF that conditions on averaged DINOv2 features from front and back views.
% in which both the front and back views of the garments are fed into the model by averaging their DINOv2 features. 
\begin{figure*}[th]
  \includegraphics[width=\textwidth]{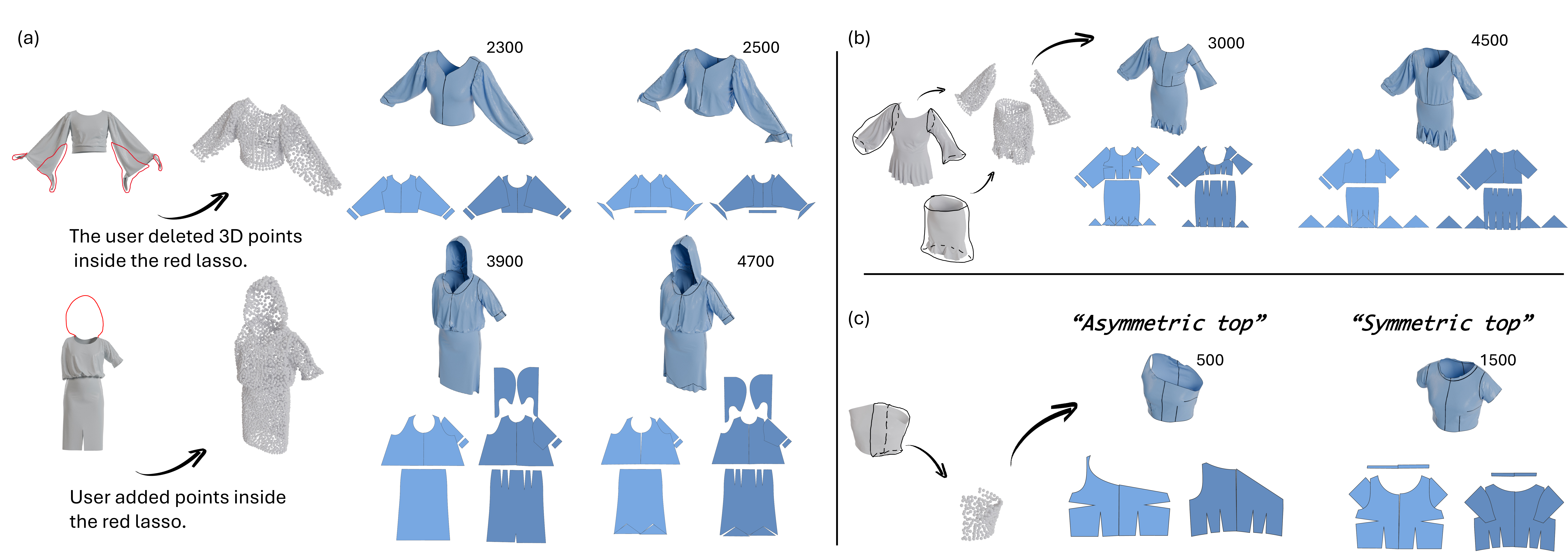}
  \caption{\textbf{Point-Cloud-Conditioned Garment Generation.} We demonstrate various point-cloud-based garment editing applications enabled by GPF. (a) illustrates how users can directly edit an existing 3D garment to guide its generation. Addition and deletion of points are achieved using our 3D interface. (b) shows garment mixing, where components of two existing 3D garments are combined to generate a new garment. (c) shows text-conditioned generation given an incomplete 3D garment. The numbers indicate the number of garment particles used to generate each sample.}
  \label{fig:point_gen}
\end{figure*}

\begin{figure*}[h]
   \includegraphics[width=\textwidth, trim={0 0 0 5cm},clip]{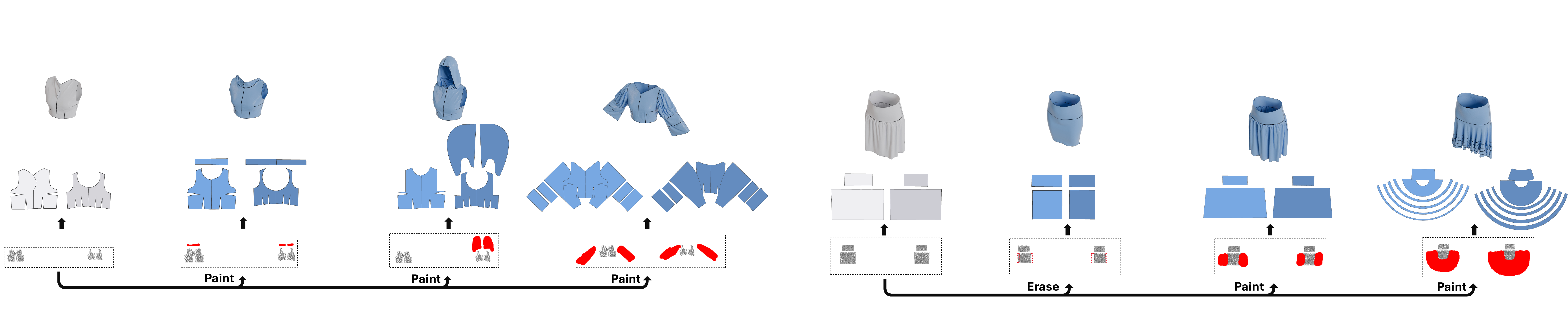}
  \caption{\textbf{Sewing Pattern Editing.} Given a generated garment shown in grey, we edit the sewing pattern and use it to guide the garment generation process. The red part illustrates the user's addition with our 2D user interface.}
  \label{fig:pattern_gen}
\end{figure*}
\begin{figure*}[h]
   \includegraphics[width=\textwidth, trim={0 0 0 4cm},clip]{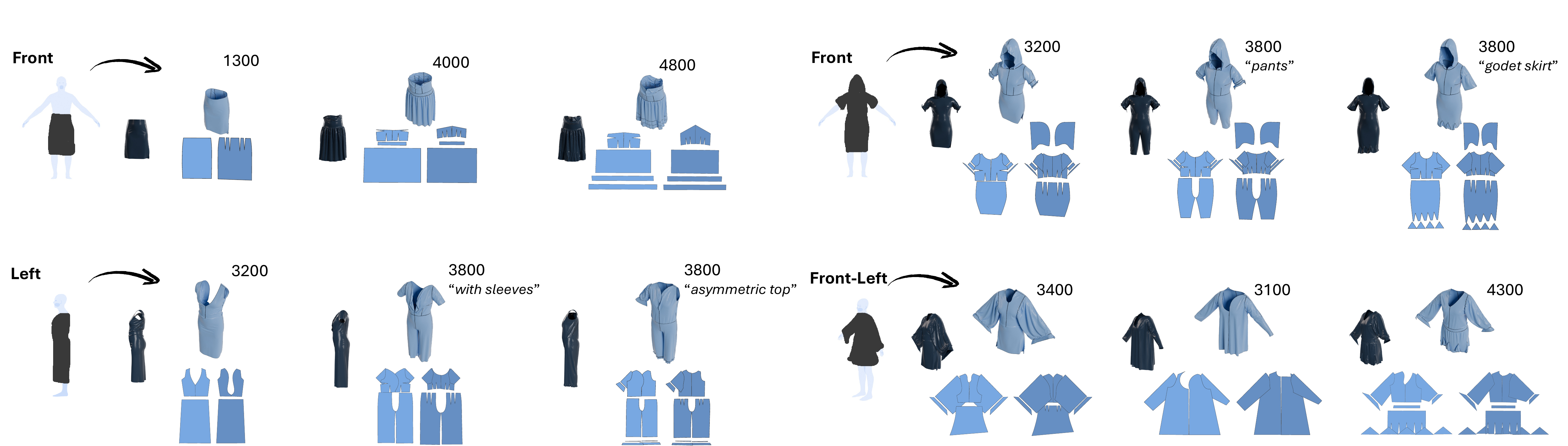}
  \caption{\textbf{Silhouette Conditioned Garment Generation.} The user paints 2D
projection to guide the garment generation process using our 2D user interface. The user can control the complexity of the generated garments as the number of points changes. The numbers indicate the number of garment particles used.
}
  \label{fig:silhouette_gen}
\end{figure*}
\paragraph{Evaluation Metrics.}
We evaluate the quality of the generated sewing patterns against the ground truth using sewing pattern metrics used in prior work~\cite{nakayama2025aipparel, li2025garmagenet, liu2023sewformer}. See the Supplementary for details. 

\begin{figure}[h] %trying to force figs apart
\centering
\includegraphics[width=\linewidth]{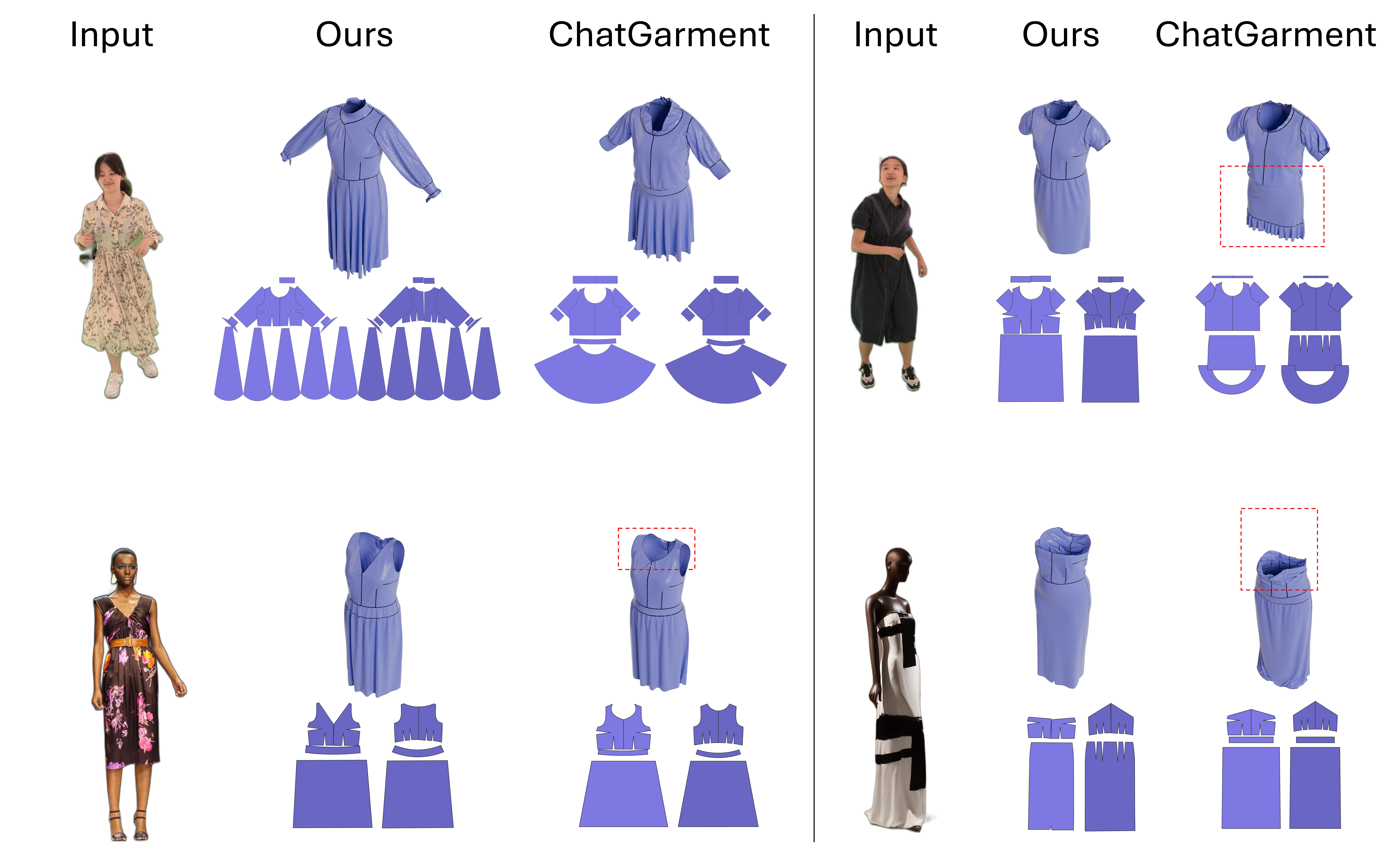}
\caption{\textbf{In-the-wild Image-conditioned Garment Generation.} Our method can generate more plausible sewing patterns than ChatGarment that match the garment style displayed in the in-the-wild images.}
\label{fig:in_the_wild}
\end{figure}

\paragraph{Results.}
\autoref{tab:image_gen} presents quantitative comparisons between our method and baselines on the GCDv2 and garment sketches datasets. For fairness, we evaluate each baseline only on the dataset it was trained on. Our method consistently outperforms all baselines across metrics except for SSR. This indicates that our method recovers more accurate sewing patterns and produces draped garments that closely match the ground truth. While program-generation-based methods (ChatGarment and D2GC) achieve slightly higher SSR, our method still achieves around $90\%$ SSR and outperforms SVG-generation-based methods (AIpparel and SewingLDM) by around $10\%$. 
% This suggests that our two-stage approach is more effective than directly predicting vectorized sewing patterns from image-conditioned data. 
% Additionally, multiview conditioning consistently improves geometry reconstruction metrics, indicating that additional images provide useful constrains for generation.
% toward the ground truth, demonstrating the effectiveness of our fine-tuning scheme across different modalities. 
We show qualitative comparisons on GCDv2 and garment sketches (\autoref{fig:image_gen}) and in-the-wild images (\autoref{fig:in_the_wild}). As shown by the red boxes, Baseline methods sometimes miss fine details such as sleeve cuffs or collars, or produce different garment styles in their generated patterns. In contrast, our method accurately captures garment style and pattern details.
% yielding a garment geometry that closely matches the ground truth. 
% \autoref{fig:in_the_wild} shows qualititave results on in-the-wild images. 

% \subsection{Ablation on Particle-to-Pattern Flow}

\section{Garment Editing Results}
% Garment Particles supports various garment-editing applications.
% that utilize the techniques described in~\autoref{sec:editing_method}. 
% We showcase these applications in this section. 

\subsection{Garment Interpolation}
\begin{figure}[h]
    \centering
   \includegraphics[width=\linewidth, trim={0 0 0 0cm},clip]{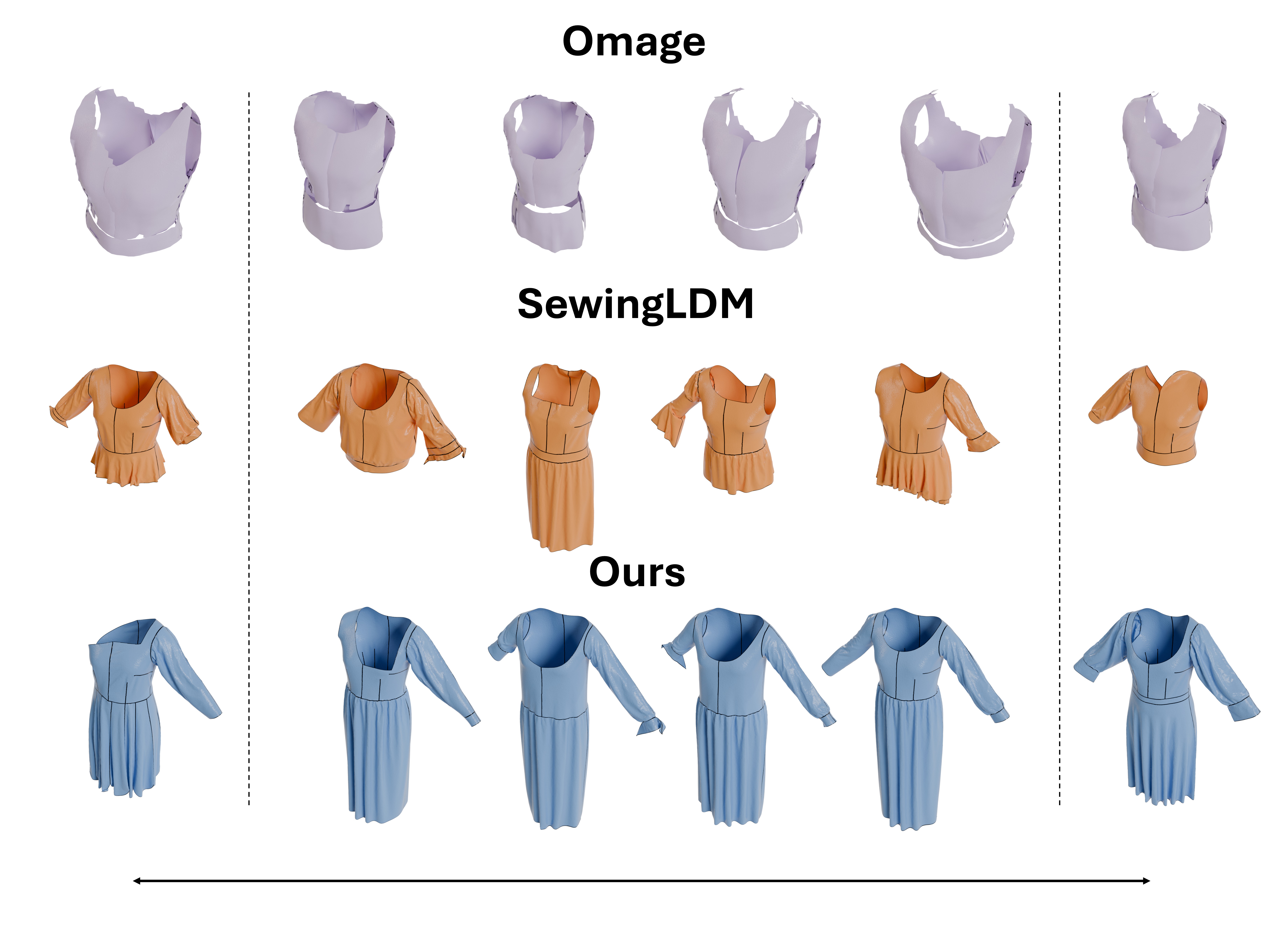}
  \caption{\textbf{Garment Interpolation Comparison.} The interpolation results of baselines exhibit non-intuitive transitions and abrupt style changes. Our results show smooth variations in the 3D geometry that transition between different garment styles.
  }
  \label{fig:interpolation}
\end{figure}

We showcase garment interpolation in~\autoref{fig:interpolation} using unconditionally generated garments from SewingLDM, Omage, and GPF. 
SewingLDM exhibits abrupt transitions due to the latent space of its vectorized sewing patterns.
% , its interpolation results lack geometric smoothness. This results in abrupt and unintuitive changes in the intermediate steps. 
Omage's geometry-image-based latent space enables smoother overall shape interpolation but lacks panel-level correspondences, \eg~the waistband varies arbitrarily in size during the transition.
% indicating that the interpolation is applied only to the overall geometry and ignores panel-level correspondences. 
In contrast, our method learn the prior over the joint 2D--3D space, enabling smooth and meaningful transitions even across topological changes, \eg~from asymmetric sleeves to symmetric ones. 
% These features make it easier for users to explore variations in garment design across the generated options from our method. 

\begin{figure*}[ht]
    \centering
    \includegraphics[width=\textwidth]{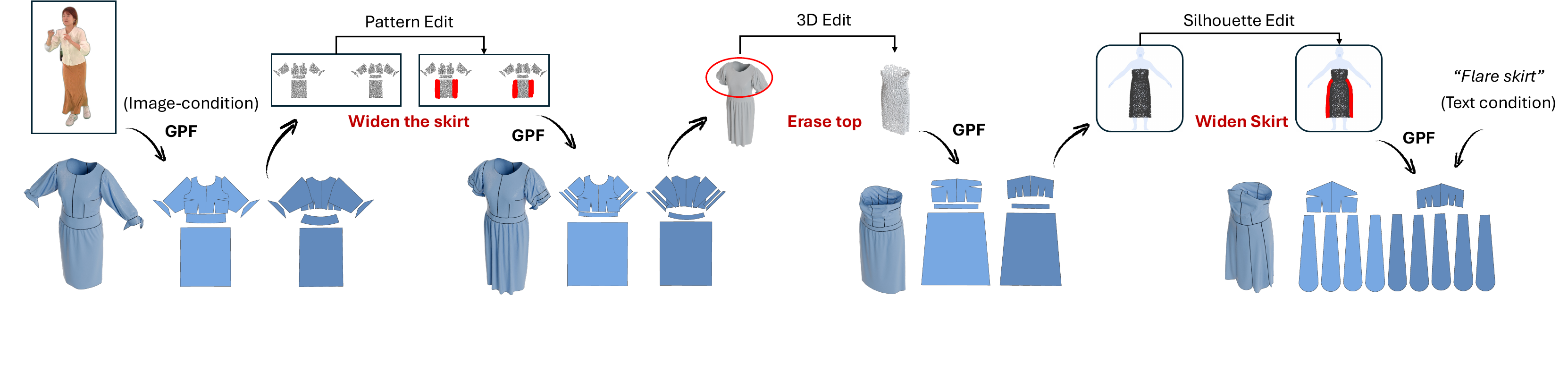}
  \caption{\textbf{Multi-step Garment Editing Session.} We show a garment editing sequence combining various editing methods enabled by GPF.}
  \label{fig:combine}
\end{figure*}

\subsection{Objective Guided Garment Editing}
% Garment particles enable objective-guided garment editing, thanks to their bidirectional 2D-3D representation of sewing patterns and garment geometry. Using the diffusion posterior sampling algorithm described in~\autoref{sec:dps}, we showcase editing results using different objectives. 
\subsubsection{Point-cloud-conditioned Sewing Pattern Generation}
% \ichao{I assume this correspond to ``3D to Sewing Pattern Generation'' and ``Unconstrained Garment Completion''?}
DPS enables garment generation from point clouds without additional training. 
% In~\autoref{fig:point_gen}, we visualize different use cases of this tool. 
In~\autoref{fig:point_gen}a, we present the generated garment assets from GPF edited directly in 3D. We generate two sewing pattern variations for each edited garment with varying particle count at input. 
% \highlight{We also include garments generated by a baseline, which is a point-cloud-to-pattern generative model. We see that while the baseline and ours both produce a suitable shirt for the top-row example, the baseline could not handle the asymmetric skirt in the bottom row because it is rarely observed in the dataset. Our model using DPS, however, was able to recover a variation of sewing patterns suitable for the point-cloud.}
% we can control the complexity of the generated patterns. 
In~\autoref{fig:point_gen}b, we show an application for mixing the styles of two garments by combining components (the dress sleeve and the godet skirt) of existing garments to create an incomplete observation $\bm{Y}$ and obtain the resulting garment using ~\autoref{sec:garment_completion}. 
% \ichao{What do you mean ``we use the setting from Y?''}
% We use the setting in~\autoref{sec:garment_completion}. 
Similarly, we observe different geometries and sewing patterns across runs and input point counts, all of which closely match the observation. We further demonstrate the ability to control generated sewing patterns with text prompts in~\autoref{fig:point_gen}c. We take a garment part (shown as a point cloud) and complete it using additional text prompts ``asymmetric'' and ``symmetric top''. Our model generates a garment that matches the input observation while obeying the specified garment style.
% demonstrating the different levels of controllability in garment editing enabled by GPF. 

\subsubsection{Sewing Pattern Editing}
\autoref{fig:pattern_gen} shows two examples where patterns generated by our method (grey) are coarsely modified (red part). The modified patterns are then used as observations for DPS. The blue garments show the generation results. We observe that all the patterns align with the inputs' designs while remaining realistic. The 2-panel skirt example shows how to convert it into narrower, wider, and circular variations, which is difficult to specify in 3D. 

\subsubsection{Silhouette-conditioned Garment Generation}
\autoref{fig:silhouette_gen} shows silhouette-conditioned garment generation with DPS from different viewpoints. For each example, the left shows the input silhouette drawings. 
% The top row shows silhouettes from the front, and the bottom row shows silhouettes from the left and front-left sides of the human body. 
We generate garment variations utilizing different particle counts and text prompts. The generations achieve a balance between alignment with the silhouette, text prompt, and garment realism. 

% ichao: I did not see any figure for this so I remove it out for now.
We showcase the fidelity and diversity trade-off by adjusting the parameter \texttt{opt\_n} in~\autoref{fig:dps_illustration}, right. Lower \texttt{opt\_n} yields more diverse output, at the expense of fidelity to the input. On the other hand, a higher value of \texttt{opt\_n} generates faithful samples, but the optimization constrains the sampling to more unimodal samples. 
% \begin{figure*}[h]
%     \centering
%     \includegraphics[width=\textwidth]{figures/multistep_edit_v4.pdf}
%   \caption{\textbf{Multi-step Garment Editing Session.} We show a garment editing sequence combining various editing methods enabled by GPF.}
%   \label{fig:combine}
% \end{figure*}
\subsubsection{Multi-step Garment Editing Session with GPF}
\autoref{fig:combine} illustrates a multi-step garment editing session using different tools from our model. Given an in-the-wild image, we obtain an initial garment using GPF. 
% from 4DDress to generate 
We edit its 2D sewing pattern to enlarge the skirt, creating wrinkles in the garment geometry that are difficult to specify in 3D. Next, we use only the bottom part of the garment for 3D conditioned garment generation. Finally, we edited the front silhouette to enlarge the skirt and generate the final garment. 
We further constrain the generated garment style with a text prompt.  

\begin{figure}[t]
\centering
\includegraphics[width=0.8\linewidth]{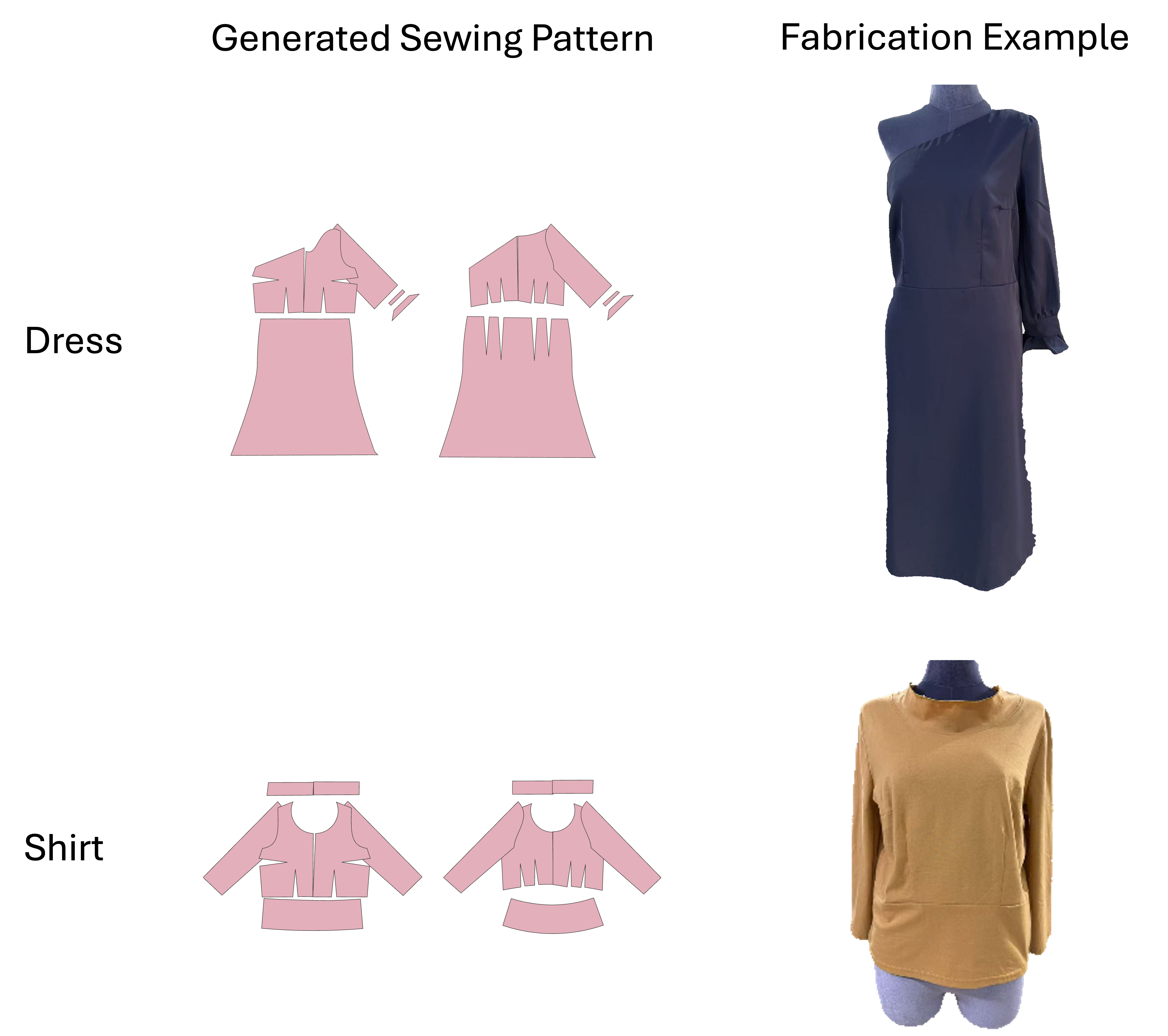}
\caption{\textbf{Fabrication Examples.} We fabricated generated sewing patterns.
% generated from our pipeline.
}
\label{fig:fabrication}
\end{figure}
\subsubsection{Fabrication Results}
We hire a tailor and fabricate two garments, an asymmetric pencil skirt and a turtle-necked shirt, generated unconditionally from our pipeline in~\autoref{fig:fabrication}. 
% Specifically, we use GPF to generate unconditionally, resulting in an asymmetric pencil skirt and a turtle-necked shirt. 
% These fabricated the sewing patterns with the help of a tailor. 
% While we needed to make minor adjustments to parts of the patterns (\eg, the sleeve cuff of the dress) and add seam allowance, the overall pattern design can be directly fabricated, resulting in wearable clothes.

\section{Limitations and Future Work}
Garment particles is a 5D point cloud representation that is a discrete sampling of a 3D surface and a 2D sewing pattern; therefore, it is difficult to represent continuous 3D surfaces and precise 2D pattern boundaries within a single-stage pipeline. The model is not suitable for fine-grained adjustments, such as the size of a dart, due to the limited particle resolution. In addition, our GPF requires the number of points as input, which may necessitate an additional mechanism, such as predicting an appropriate number of points for a given query. During iterative editing, the garment particles are resampled from the model, so fine details are not precisely preserved.

Editing via DPS is time-consuming, as users must wait before seeing results. For interactive applications, it is desirable to instantly obtain results to support direct manipulation, such as updates during mouse dragging. In addition, a garment defined by a 2D pattern can exhibit different 3D geometries depending on body size, posture, and fabric properties. Extending our 5D model to account for such variations is a promising direction for future work. Lastly, our model is trained only on garments produced using GarmentCode~\cite{GarmentCode2023}. We aim to expand our training dataset to include a wider variety of garments, such as those from~\cite{li2025garmagenet}. 

Finally, we cannot guarantee that the generated sewing pattern exactly matches the input garment particles since PPF is purely data-driven. Future work includes enforcing hard constraints, such as preventing interior particles from leaving the regions of 2D panels.
\section{Conclusion}
We present \methodname, a 2D--3D symmetric garment representation that jointly encodes the sewing pattern and its draped garment geometry as a 5D point cloud. Using garment particles, we train Garment Particles Flow (GPF), a flow-based generative framework that learns a semantically rich prior space that enables state-of-the-art garment generation. More importantly, GPF naturally supports garment editing applications through diffusion posterior sampling with various objectives, including sewing pattern editing, as well as point-cloud- and silhouette-conditioned garment generation.
% that enables both casual garment generation and various garment editing applications.
% via diffusion posterior sampling. 
To recover simulation-ready sewing patterns, we propose Particles-to-Pattern flow to convert the garment particles to vectorized sewing patterns. 
\begin{comment}
GPF learns a semantically rich prior space that enables state-of-the-art garment generation. More importantly, GPF naturally supports garment editing applications through diffusion posterior sampling with various objectives, including garment interpolation, sewing pattern editing, and point-cloud- and silhouette-conditioned garment generation.
\end{comment}

% via diffusion posterior sampling with various objectives. 
% These advancements unlock a new set of tools for garment design, where both casual garment generation and intuitive garment editing are essential. 
\begin{comment}
We believe that our work is a significant step forward, towards making generative AI more practical and integrated into garment designers' daily workflow, bridging the gap between creative intent and technical execution for pattern making.  
\end{comment}

% this will show up only in the final version
\begin{acks}
% We thank ooxx. 
This research is supported by JST ASPIRE, JPMJAP2401, Initiative on Recommendation Program for Young Researchers and Woman Researchers, Information Technology Center, The University of Tokyo, LVMH, Google, and the National Science Foundation Graduate Research Fellowship Program. 
\end{acks}

\bibliographystyle{ACM-Reference-Format}
\bibliography{bibliography}

\clearpage
% \title{Supplementary Material for Garment Particles: A 2D–3D Symmetric Garment Representation for Generation and Editing}
% \maketitle
% \input{sections/figure_only_page}
\appendix
\twocolumn[
\begin{center}
    {\textbf{Supplementary Material for Garment Particles: A 2D–3D Symmetric Garment Representation for Generation and Editing}}
    \vspace{1em}
\end{center}
]
\section*{Table of Contents}
\startcontents
\printcontents{ }{1}{}
\section{Implementation, Dataset, and Metrics Details}
\subsection{Garment Particles Construction Details}
We construct our garment particles datasets on  GarmentCodeDatav2 (GCDv2)~\cite{korosteleva2024garmentcodedatadataset3dmadetomeasure}. Specifically, we first separate the front and back panels using the provided panel names. The front and panel panels are packed using the algorithm outlined in~\autoref{sec:garment_particle_repr}. To resolve panel overlaps, we use an iterative algorithm that pushes overlapping panels apart based on the distance and direction between their centers. Additionally, to ensure greater consistency across garments, we define a tree structure over the panels based on their semantic relationships. The repulsion algorithm is then applied first between siblings and then between children and parents. To ensure robustness, we pad each panel by 5 centimeters around its boundary when checking for overlaps. The repulsion algorithm will run until there are no overlaps between any panels or it reaches 500 steps. After constructing the garment particles, we filter out data samples whose sewing pattern $U$ exceeds the bounding box $[-150, 150] \times [-80, 220]$. In total, we obtain ~120k valid examples out of 130k garments in GCDv2.  

\subsection{GPF Training Details}
To train the text-conditioned GPF model, we use 32 $\times$ NVIDIA H100 GPUs for a total of $210{,}000$ iterations. 
To speed up training, we use Pytorch FSDP2~\cite{zhang2024simplefsdpsimplerfullysharded} and Flash Attention~\cite{dao2022flashattentionfastmemoryefficientexact}. We use a dynamic batch sampler that distributes roughly equal numbers of tokens across GPUs, resulting in an average batch size of 250. We use a learning rate of 0.0001 with gradient clipping at 1. The training takes around 1.5 days. 

\subsection{Text Caption Dataset Construction}
We construct a new text caption dataset to train our text-conditioned GPF model. We procedurally generate the text captions from the design parameters given in the GCDv2 dataset. Each text prompt consists of a set of short, keyword phrases, describing the make of different components of the garments (\eg~``with'' v.s. ``without sleeves'', ``fitted'' v.s. ``loose shirt'', ``pants'' v.s. ``skirts'', etc.). Compared with existing text-prompt datasets, such as GCD-MM~\cite{nakayama2025aipparel}, our curated caption focuses more on garment style and is therefore better suited for integration with garment editing applications. During training, we sample a subset of these keywords and combine them into a single input for the GPF model. 

\subsection{Particles-to-Pattern Flow Training Details}
We use the LightningDiT-L variant as our Flow architecture, which consists of around $500M$ trainable parameters. We apply both panel and edge-level positional embedding during training. In notations, we have 
\begin{equation}
    \bm{v}_{\varphi}(\mathcal{P}_t, t; \bm{X}) = \operatorname{DiT}\paren{\operatorname{Flatten}(\bm{W}_{pos}+ \mathcal{P}_t), t; \bm{X}}
\end{equation}
where $\bm{W}_{pos} = \bm{W}_{panel} \oplus \bm{W}_{edge}$ is the outer sum of panel embeddings and edge embeddings, and the $\operatorname{Flatten}$ operator flattens the panel and edge dimension. We additionally apply RoPE~\cite{su2024roformer} to enhance the positional information. We use cross-attention to inform the network about $\bm{X}$. 
We use the same training hyperparameters as GPF model training on 16$\times$ NVIDIA H100 GPUs for a total of 160k iterations. 
\subsection{Objective Guided Editing Hyperparameters}
We found that different sets of hyperparameters, such as the sampling steps $T$, \texttt{stop\_t}, learning rate $\eta$, and optimization steps \texttt{opt\_n}, work well for different objectives and applications. However, in general, for EMD loss, we use $\eta \in [0.02, 0.1]$, $\texttt{opt\_n} \in\set{1, 2, 3, 4}$, $T \in [250, 1000]$, and $\texttt{stop\_t} \in [0,3, 0.8]$. For Chamfer Distance as the objective, we keep the same $T$ and \texttt{stop\_t}, but increase $\texttt{opt\_n}$ to 10 and decrease $\eta$ to between $[0.01, 0.02]$.

\begin{figure}[th]
    \centering
    \includegraphics[width=\linewidth]{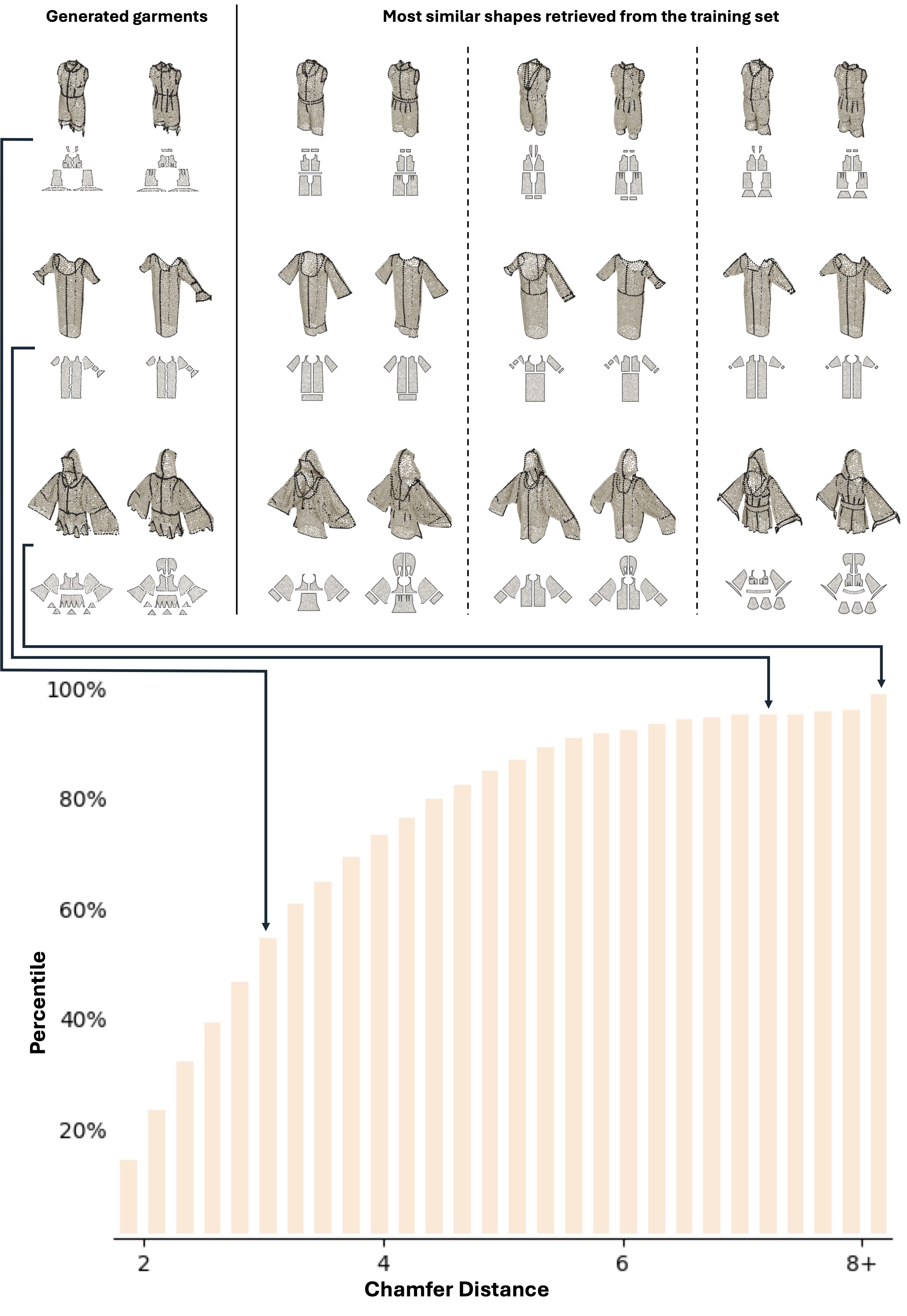}
    \caption{\textbf{Closest Query Visualization}. \textit{(Top)} We visualize the top three nearest neighbors in the training set to our generated garment particles (leftmost column). Our GPF model can generate novel garments with a distinct style compared to the training set. \textit{(Bottom)} We plot the distance of our generated sets to the training set as a cumulative plot. The arrows indicate the bins to which each of the visualized garments belongs.}
    \label{fig:shape_novelty}
\end{figure}

\subsection{Generation Metrics}
We use a different set of metrics to measure the diversity and realism of our generated garments relative to the ground-truth garments in the test set. Specifically, they measure the following. 
\begin{enumerate}
    \item \textit{Generation diversity and distribution} using the Coverage (COV) score, Minimum Matching Distance (MMD), and 1-Nearest Neighbor classification Accuracy (1-NNA).
    % , which are standard distribution metrics in 3D. 
    Following~\cite{zhang2024clay, li2025garmagenet}, we also report pointcloud-FID (p-FID) to assess feature-embedding similarity against the reference set. 
    % Ideally, the generated set should obtain high COV, a 1-NNA close to $50\%$, and low MMD, p-FID, and p-KID. 
    \item \textit{Draping quality} using simulation-success rate (SSR) of the generated garments. For all our baselines, we use GCDv2's provided draping simulator. 
    \item \textit{The alignment between generated garment and input text prompts} using CLIPScore~\cite{Hessel2021CLIPScoreAR}. 
    % A higher CLIPScore indicates a better alignment with the text prompt. 
\end{enumerate}
We define each of the metrics below. 
\paragraph{Coverage Score (COV)}
Coverage Score measures the percentage of reference garments that are matched by at least one generated sample. Given the set of generated garments $S_g$ and reference garments $S_r$, we compute the score as
\begin{equation}
    \text{COV}(S_g, S_r) = \frac{100}{|S_r|} \left| \left\{ \arg\min_{y \in S_r} D(x, y) \mid x \in S_g \right\} \right|,
\end{equation}
where $D(\cdot, \cdot)$ denotes the Chamfer distance between two point clouds. A higher score indicates a more diverse set of generated samples that better covers the reference distribution. We uniformly sample 8,192 points on each garment mesh surface for computation.

\paragraph{Minimum Matching Distance (MMD)}
Minimum Matching Distance measures the fidelity of the generated set by computing the average distance from each reference sample to its closest generated sample. Given the set of generated garments $S_g$ and reference garments $S_r$, we compute the score as
\begin{equation}
    \text{MMD}(S_g, S_r) = \frac{1}{|S_r|} \sum_{y \in S_r} \min_{x \in S_g} D(x, y),
\end{equation}
where $D(\cdot, \cdot)$ denotes the Chamfer distance between two point clouds. A lower MMD indicates that the generated samples are closer to the reference distribution. We uniformly sample 8,192 points on each garment mesh surface for computation.

\paragraph{1-Nearest Neighbor Accuracy (1-NNA)}
The 1-Nearest Neighbor Accuracy evaluates whether the generated and reference samples are distinguishable by a nearest neighbor classifier. For each sample in $S_g \cup S_r$, we find its nearest neighbor (excluding itself) and check whether they belong to the same set:
\begin{equation}
    \text{1-NNA}(S_g, S_r) = 100 \times \frac{\sum_{x \in S_g} \mathbf{1}[\mathrm{NN}(x) \in S_g] + \sum_{y \in S_r} \mathbf{1}[\mathrm{NN}(y) \in S_r]}{|S_g| + |S_r|},
\end{equation}
where $\mathrm{NN}(\cdot)$ returns the nearest neighbor in $S_g \cup S_r$ using Chamfer distance. An ideal generative model produces samples indistinguishable from the reference set, yielding a 1-NNA close to $50\%$.

\paragraph{Pointcloud FID (p-FID)}
Pointcloud FID measures the similarity between the feature distributions of generated and reference point clouds. We first extract features using a pretrained point cloud encoder, then compute the Fréchet distance between the two Gaussian-fitted distributions:
\begin{equation}
    \text{p-FID} = \|\boldsymbol{\mu}_r - \boldsymbol{\mu}_g\|_2^2 + \mathrm{Tr}\left( \boldsymbol{\Sigma}_r + \boldsymbol{\Sigma}_g - 2\left( \boldsymbol{\Sigma}_r \boldsymbol{\Sigma}_g \right)^{1/2} \right),
\end{equation}
where $(\boldsymbol{\mu}_g, \boldsymbol{\Sigma}_g)$ and $(\boldsymbol{\mu}_r, \boldsymbol{\Sigma}_r)$ are the mean and covariance of the generated and reference feature embeddings, respectively. A lower p-FID indicates greater similarity to the reference distribution.

\paragraph{Simulation Success Rate (SSR)}
Simulation Success Rate measures the physical plausibility of generated garments by evaluating whether they can be successfully draped on a human body without simulation failures (\eg, mesh interpenetration or divergence):
\begin{equation}
    \text{SSR} = \frac{N_{\text{success}}}{N_{\text{total}}} \times 100,
\end{equation}
where $N_{\text{success}}$ is the number of garments that complete the draping simulation without errors, and $N_{\text{total}}$ is the total number of generated garments. A higher SSR indicates that the model generates more physically valid garment geometries.

\paragraph{CLIPScore}
CLIPScore measures the semantic alignment between generated garments and input text prompts. For each successfully draped garment, we render $|V|=20$ views uniformly distributed along the equator and compute the average cosine similarity between image and text embeddings:
\begin{equation}
    \text{CLIPScore} = \frac{100}{|V|} \sum_{v \in V} \cos\left( \mathbf{E}_I(I_v),\, \mathbf{E}_T(t) \right),
\end{equation}
where $\mathbf{E}_I$ and $\mathbf{E}_T$ are the CLIP image and text encoders, $I_v$ is the rendered image from view $v$, and $t$ is the input text prompt. A higher CLIPScore indicates better alignment between text and garment.

\subsection{Sewing Pattern Metrics}\label{sec:pattern_metrics}
To evaluate the quality of the generated sewing pattern relative to the ground truth, we use the sewing pattern metrics defined in prior work~\cite{nakayama2025aipparel, li2025garmagenet}. Specifically, we use (1) panel accuracy (Panel Acc): the percentage of garments with the correct number of panels, (2) panel-wise intersection-over-union (Panel IOU): the average IOU between the generated and ground-truth panels, and (4) stitch accuracy (Stitch Acc): the percentage of correctly predicted stitching pairs. We also measure the simulation success rate (SSR) and the 3D Chamfer Distance (CD) of the draped garments. 

\subsection{Extended Discussion on Limitations and Future Works}
\paragraph{Garment Particle Flow}
While our GPF module learned the implicit consistency between the 2D sewing pattern 3D garment geometry from data, we do not gurantee hard constraints on this front. For example, manufaturing constraints such as developability of the 3D surface and near-isometry constraint between the pattern and the draped garment geometry. Enforcing these constraints either directly during training, or via inference-time scaling are interesting future work directions.

Additionally, our GPF model is trained on GarmentCodeData. While being the largest sewing pattern dataset online, they are synthetic and lack important components such as pockets, frills. Another future work direction is to integrate more realistic sewing pattern datasets (e.g., GarmageSet~\cite{li2025garmagenet}) or potentially in-the-wild sewing patterns to the training pipeline.

Lastly, our GPF model is only trained on a single body type with the same pose. This limits our applications to be performed on the same human. One could extend our GPF to handle multi human body + human pose input, and extending our model for garment refitting applications. 
\paragraph{Particle-to-Pattern Flow}
Similar to GPF, our PPF model also learns the sewing pattern reconstruction purely from data, with a conditional generative model. While this approach make PPF more robust to noisy garment particle inputs, we cannot guarantee that the reconstructed garment is strictly consistency with respect to the input. One promising future direction is the improve the consistency via post-training strategies such as Flow-GRPO~\cite{liu2025flow}, with non-differentiable rewards such as IOUs and accuracies as rewards. 

Additionally, our sewing pattern representation still use the format popularized by GarmentCodeData~\cite{korosteleva2024garmentcodedatadataset3dmadetomeasure}, which uses one-to-one stiches and rigid transformation based panel initializations. Recent works such as GarmageNet extended the representation to use point-to-point stitching and point-wise panel initialization, enabling more complext garment modeling capabilities. It would be interesting to extend our PPF module to allow these more flexible sewing pattern representations, therefore enabling more complext garment generation and editing. 
\section{User Interfaces}
To facilitate the use of GPF for various editing tasks in both 2D and 3D design spaces, we developed two interfaces as follows. 

\begin{figure}[ht]
    \centering
    \includegraphics[width=\linewidth]{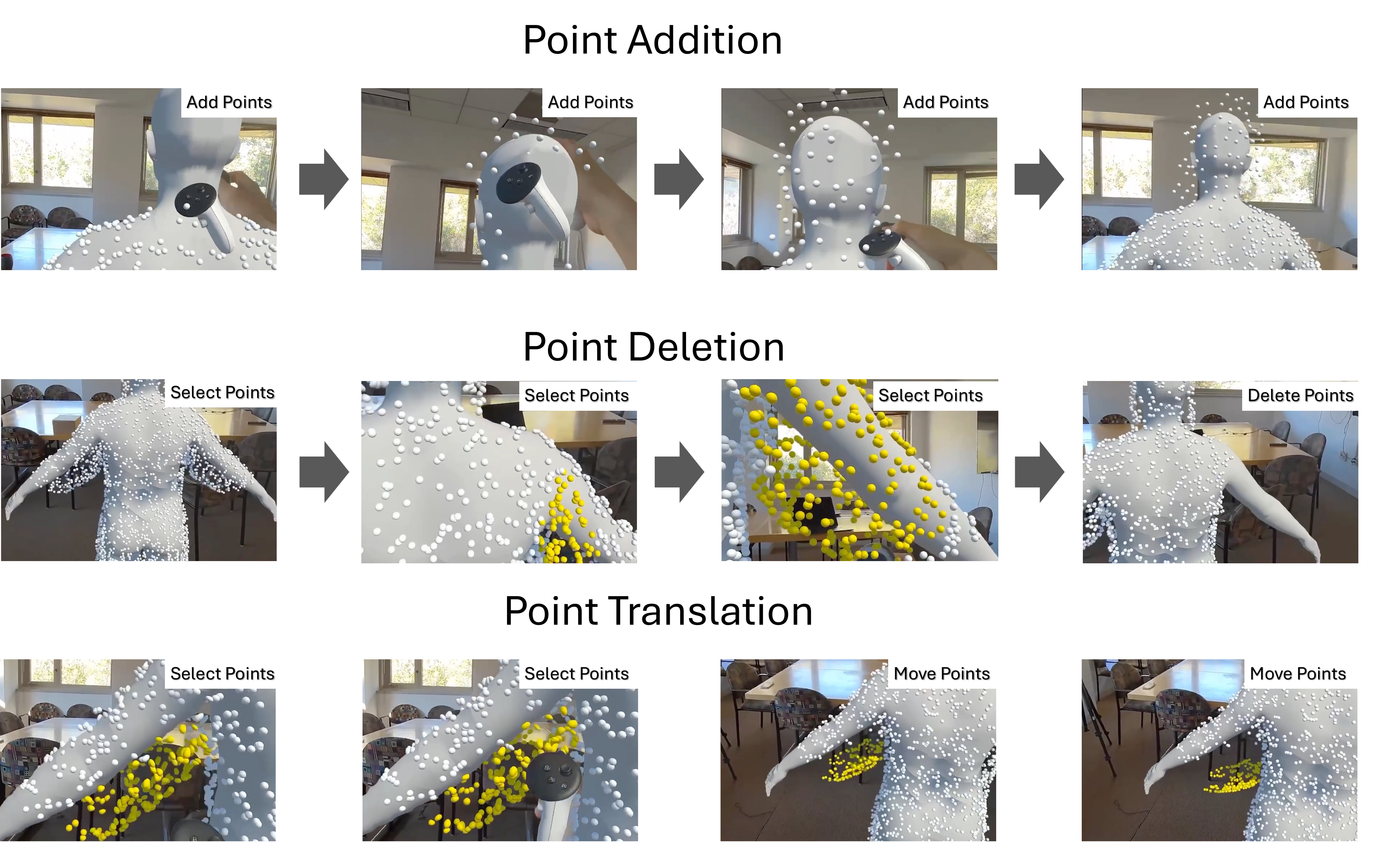}
    \caption{\textbf{3D Interface Illustration}. Our 3D interface allows users to directly manipulate 3D geometry with operations such as point addition, deletion, and translation.}
    \label{fig:3d_demo}
\end{figure}
\subsection{3D Interface}
% We developed a 3D user interface that displays the 3D garment geometry as a point cloud in a virtual reality setting. The garment geometry is represented as a point cloud, and a set of editing tools is developed to allow users to casually edit it with a VR controller. Specifically, we implemented point addition, deletion, and translation. The 3D environment is illustrated in~\autoref{fig:3d_demo}.
We developed a 3D user interface for interactive garment editing in augmented reality (AR), in which garment geometry is represented as a point cloud. Within the immersive environment, we designed and implemented a set of editing tools that enable users to directly manipulate the point cloud with controllers, including point addition, deletion, and translation.
The system was implemented in \emph{Unity}, a cross-platform 3D game engine, using the built-in render pipeline. \emph{Meta Quest 3} was leveraged as the head-mounted display (HMD) to present the AR environment and support user interaction during garment editing. An overview of the 3D interactions is illustrated in~\autoref{fig:3d_demo}.
\paragraph{Addition}
Users can add points by pressing the \emph{A} button on the right controller, which inserts new points in the neighborhood of the controller’s current position.
\paragraph{Deletion}
To delete points, users first press the \emph{Index Trigger} on the right controller to select points near its position, which are visually highlighted in yellow. Pressing the \emph{B} button subsequently removes the selected points.
\paragraph{Translation}
Point translation follows a procedure similar to deletion. Users first select the target points and then manipulate the \emph{Joystick} on the right controller to move the selected points in a non-linear manner.

\begin{figure}[ht]
    \centering
    \includegraphics[width=\linewidth]{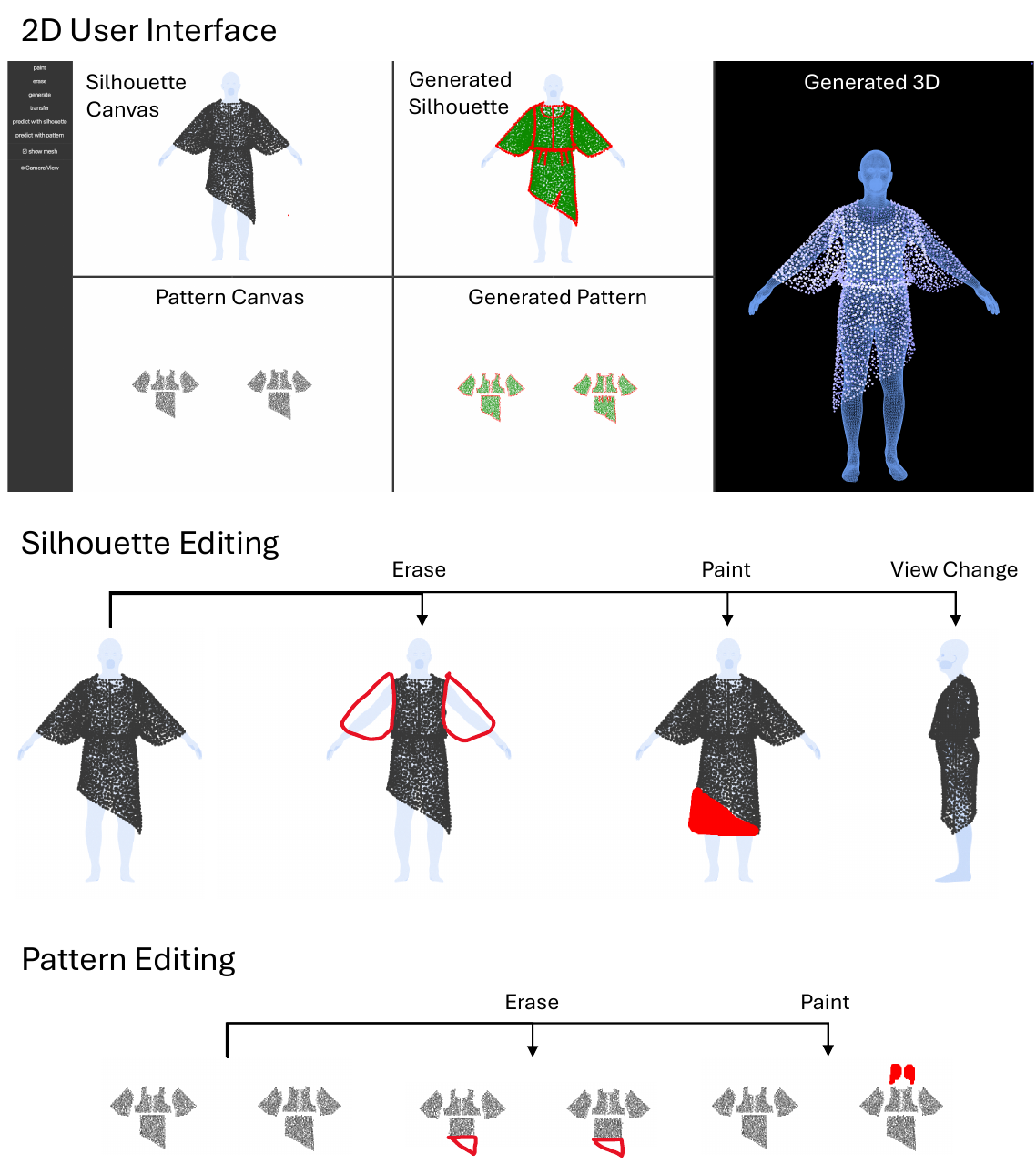}
    \caption{\textbf{2D Interface Illustration}. Our 2D interface supports silhouette and pattern editing with paintbrushes. The top row shows our interface's layout, the next two rows show different operations we allow for silhouette and pattern editing.}
    \label{fig:2d_demo}
\end{figure}
\subsection{2D Interface}
% We implemented a 2D user interface that supports direct editing of sewing patterns and silhouettes from arbitrary camera angles. The interface, which is illustrated in~\autoref{fig:2d_demo}. We implement the editing operations as a canvas tool that allows users to paint or erase on both the silhouette and pattern canvases. The edited canvas is then converted to a 2D point cloud by sampling the painted area and sent to GPF for various guided generation tasks. 
We implemented a 2D user interface that supports direct editing of sewing patterns and silhouettes from arbitrary camera viewpoints, as illustrated in~\autoref{fig:2d_demo}. Editing operations are implemented through a canvas-based tool that allows users to paint or erase on both the silhouette and pattern canvases. The edited canvas is subsequently converted into a 2D point cloud via area sampling and forwarded to GPF for various guided generation tasks. 
\section{Additional Results}
\subsection{Ablation Study on Particles-to-Pattern Flow}
We conduct an ablation study to evaluate the effectiveness of our particles-to-pattern flow module relative to alternative architectural choices. Specifically, we compare our \textit{flow}-based network with two variants. \textit{Regression}-based variant uses a feedforward network to directly predict the vectorized sewing pattern given garment particles. We use the same architecture as our particles-to-pattern flow model, but simply change the loss to mean squared error. \textit{Delaunay} variant reconstructs the sewing pattern as a flat triangle mesh by performing a sequence of training-free operations. Specifically, the garment particles are first clustered into panels using DBSCAN based on their layout when projected into the domain via $\pi_D$. Then, each cluster is triangulated with a Delaunay triangulation to obtain a convex hull of each panel. Finally, the boundary for each panel is recovered by removing triangles whose three vertices have all three boundary flags set to positive (on the boundary). 

\begin{table}[th]
\centering
\begin{tabular}{l|cccc}
\toprule
Method & Panel Acc & Edge Acc & Stitch Acc & Panel IOU \\
\midrule
\multicolumn{5}{l}{\textit{No Noise}} \\ \hline
    \textit{Delaunay}     & \textbf{0.9912} & -- &  --      &        0.9034 \\
    \textit{Regression}          & 0.9902 & 0.9177 & 0.6898 & 0.6727 \\
    \textit{Flow}    & 0.9775 & \textbf{0.9430} & \textbf{0.9333} & \textbf{0.9181} \\
\midrule
\multicolumn{5}{l}{+ 1\% \textit{Noise}} \\ \hline
    \textit{Delaunay}  & 0.0947 &--  &    --    &    0.6688    \\
    \textit{Regression}       & \textbf{0.9902} & 0.9074 & 0.6829 & 0.6717 \\
    \textit{Flow} & 0.9365 & \textbf{0.8932} & \textbf{0.8770} & \textbf{0.8815} \\
\midrule
\multicolumn{5}{l}{+ 2\% \textit{Noise}} \\ \hline
    \textit{Delaunay}  & 0.0156 & -- &    --    &  0.4327      \\
    \textit{Regression}       & \textbf{0.9902} & 0.8951 & 0.6727 & 0.6687 \\
    \textit{Flow} & 0.9355 & \textbf{0.9051} & \textbf{0.8793} & \textbf{0.8822} \\
\midrule
\multicolumn{5}{l}{+ 5\% \textit{Noise}} \\ \hline
    \textit{Delaunay}  & 0.0449 &--  &    --    &      0.1666  \\
    \textit{Regression}       & \textbf{0.9658} & 0.8596 & 0.6379 & 0.6524 \\
    \textit{Flow} & 0.9355 & \textbf{0.8848} & \textbf{0.8540} & \textbf{0.8631} \\
\midrule
\multicolumn{5}{l}{+ 10\% \textit{Noise}} \\ \hline
    \textit{Delaunay}  & 0.0088 & -- &    --    & 0.1306       \\
    \textit{Regression}       & 0.7324 & 0.8187 & 0.5583 & 0.5953 \\
    \textit{Flow} & \textbf{0.8955} & \textbf{0.8834} & \textbf{0.8348} & \textbf{0.8210} \\
\bottomrule
\end{tabular}
\caption{\textbf{Ablation Study on Particle-to-Pattern Module}. We compare variants of the Particle-to-Pattern Module across different levels of noise added to the first two coordinates.}
\label{tab:ablation}
\end{table}

\begin{table*}[!t]
    \centering
    % \begin{adjustbox}{width=\textwidth, center}
    \renewcommand{\arraystretch}{0.9}% Tighter
    \begin{tabular}{l cc cc cc}
        \toprule
        \textbf{Method} & \textbf{Panel Acc} (\%) $\uparrow$& \textbf{Panel IOU} (\%) $\uparrow$ & \textbf{Edge Acc} (\%) $\uparrow$ & \textbf{Stitch Acc} (\%) $\uparrow$ & \textbf{SSR} (\%) $\uparrow$ & \textbf{CD} ($\times 10^3$)  $\downarrow$ \\
        \midrule
        \multicolumn{6}{l}{\textit{GCDv2}} \\
        \midrule
        Ours-Singleview & 83.01 & 77.54 & 78.20 & 69.47 & 89.84 & 7.0 \\
        Ours-Multiview & \textbf{85.35} & \textbf{79.63} & \textbf{80.09} & \textbf{71.49} & \textbf{90.23} & \textbf{5.5} \\
        \hline
        \multicolumn{6}{l}{\textit{Garment Sketches}} \\
        \midrule
        Ours-Singleview & 81.25 & 76.07 & 77.05 & 66.86 & \textbf{92.97 }& 8.7 \\
        Ours-Multiview & \textbf{82.71} & \textbf{78.14} & \textbf{79.01} & \textbf{67.67} & 88.18 & \textbf{8.4} \\
        \bottomrule
    \end{tabular}
    % \end{adjustbox}
    \caption{
    \textbf{Multiview-image-conditioned Garment Generation.}}
    \label{tab:multi_view_gen}
\end{table*}

% Reconstruction	Panel Acc	Edge Acc	Stitch Acc	IOU	SSR
% Ours-finetuned	0.8301	0.782	0.6947	0.7754	
% Ours	0.8818	0.8385	0.7748	0.8315	0.970703125
% ours-sketch	0.7705	0.7934	0.7182	0.7481	0.982421875
% AIpparel	0.7549	0.7036	0.6467	0.7547	0.875
% ChatGarment	0.05859375	0.4443254173	0.3772075176	0.6267241836	0.951171875
% SewingLDM	0.2060546875	0.5047988892	0.4348564744	0.544708848	0.8681640625
We compare the performance on the particles-to-pattern reconstruction task, using the sewing pattern metrics as described in~\autoref{sec:pattern_metrics}. \autoref{tab:ablation} shows the comparison with different levels of noise added to the pattern space of the particles (\ie first two coordinates), mimicking the noisy generation result from GPF's output. Because the \textit{Delaunay} variant does not predict edges and stitching information, we omit their scores for these two columns. The table shows that when no noise is added, the \textit{Delaunay} variant performs the best in terms of panel IOU and accuracy. This suggests that the classical algorithm can almost completely recover the sewing pattern shapes without precision loss \textit{if} there is no corruption in the garment particles. However, even with 1\% noise added to the pattern coordinates, its performance drops drastically, because DBSCAN (clustering) and the triangulation process are very sensitive to outliers. Comparatively, the \textit{regression}-based variant achieves better robustness against noisy input, but still is prone to error when the amount of noise added exceeds 5\%. This is because a regression-based model learns a deterministic output for each garment particles, making it susceptible to out-of-distribution data. Lastly, the \textit{Flow}-based variant retains its overall performance across all levels of added noise. This is because its formulation of the reconstruction task as a conditional generation problem, enabling the model to still output sensible patterns when the input is corrupted.

\paragraph{Effect of Boundary Flag.}
\begin{table}[th]
\centering
\begin{tabular}{l|cccc}
\toprule
Method & Panel Acc & Edge Acc & Stitch Acc & Panel IOU \\
\midrule
\textit{PPF w/o BFlag}          & 0.9121 & 0.8508 & 0.8485 & 0.9041\\
\textit{PPF}    & \textbf{0.9775} & \textbf{0.9430} & \textbf{0.9333} & \textbf{0.9181}\\
\bottomrule
\end{tabular}
\caption{\textbf{Ablation Study on Boundary Flag}.}
\label{tab:ablation_bflag}
\end{table}
\autoref{tab:ablation_bflag} shows the effect of having the boundary flag as input to the PPF module, for the sewing pattern reconstruction task. The resulting improvements in panel, edge, and stitch accuracies demonstrate boundary flag's effectiveness in recovering discrete structures from garment particles.

\subsection{Unconditional Generation}

\subsubsection{Generation Gallery}
\begin{figure*}
    \centering
    \vspace{-4em}\includegraphics[width=\textwidth]{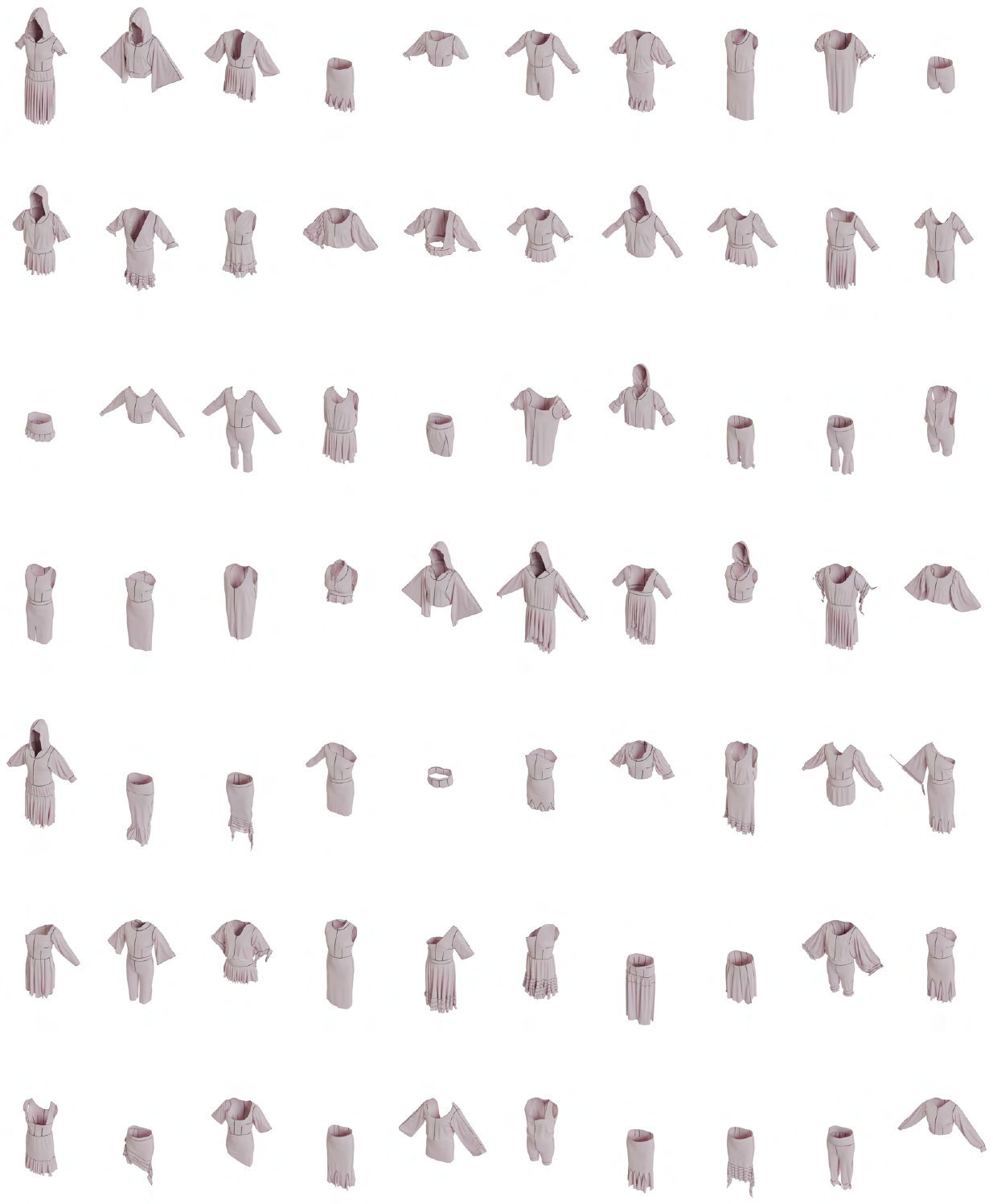}
    \vspace{-5em}
    \caption{\textbf{Unconditional Generation Gallery.}}
    \vspace{-5em}
    \label{fig:uncond_supp}
\end{figure*}
In~\autoref{fig:uncond_supp}, we showcase a gallery of our generated garments from GPF without conditioning. The generated garment particles are converted into a draping pattern using our particles-to-pattern flow module. As shown in the figure, we can generate a diverse set of garments, from simple to complex panel layouts. This demonstrates the pipeline's generation capability.
\subsubsection{Generation Novelty Analysis}
We validate our method's ability to generate novel garments not included in the training dataset. 
Following~\cite{Siddiqui_2024_CVPR}, we generate $1{,}024$ garments using our GPF model without conditioning. In \autoref{fig:shape_novelty}, we show three generated garments and their top three nearest neighbors in the training set, measured using the 3D Chamfer distance. We also report the distance of this garment set to its nearest neighbor in the training set. These results show that our model can generate novel garment styles even at the 30th-percentile level (top row), demonstrating that it learns to generate unseen combinations of sewing patterns and garment geometry.  
\subsubsection{Text-Conditioned Generation Additional Visualization}

\begin{figure*}
    \centering
    \includegraphics[height=0.97\textheight]{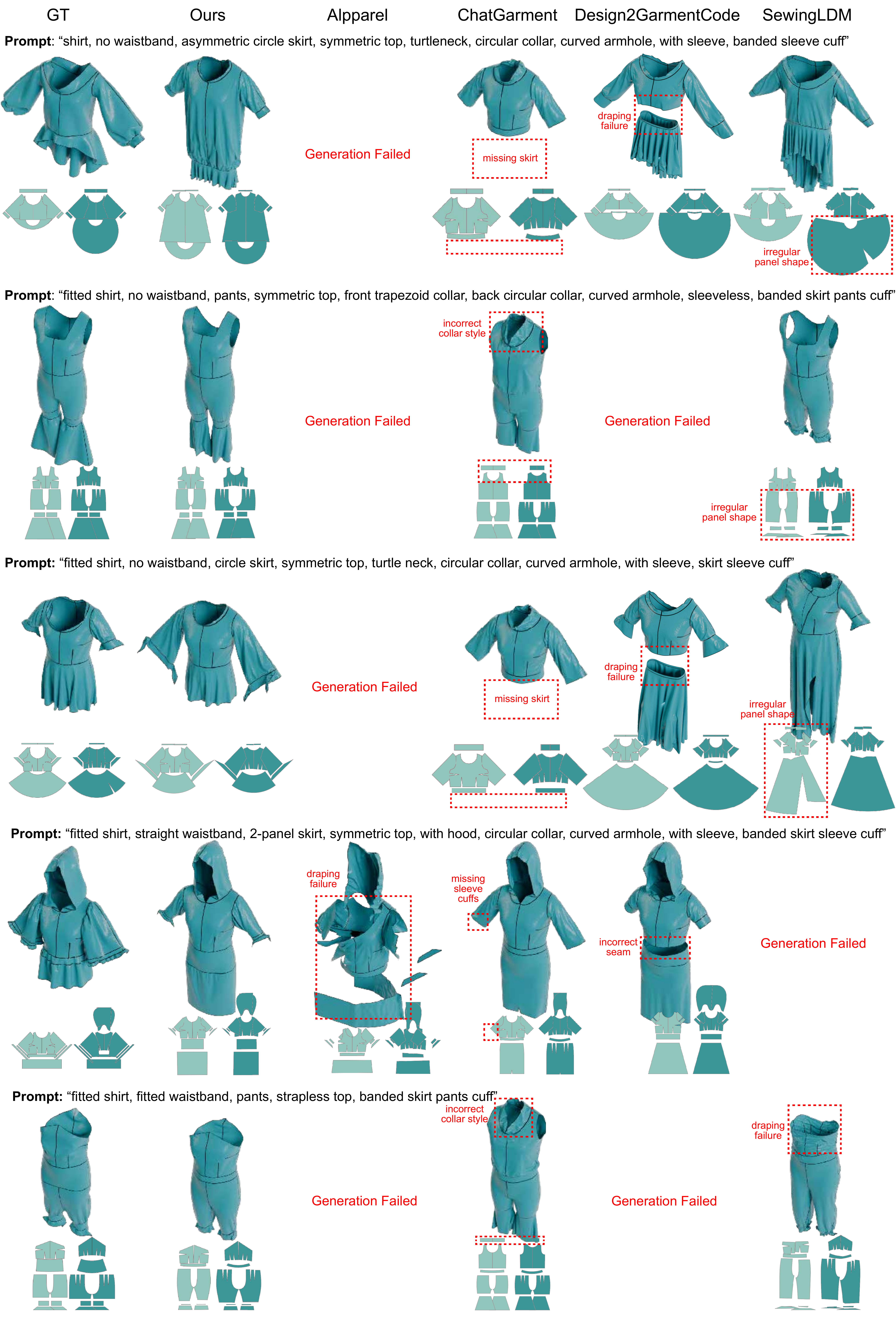}
    \vspace{-1em}
    \caption{\textbf{Text-Conditioned Generation: Additional Visualization.}}
    \label{fig:text_supp}
\end{figure*}

We showcase additional comparison for text-conditioned garment generation in~\autoref{fig:text_supp}. The baselines exhibit artifacts and even fail to produce the garment geometry, as shown by the red boxes. Compared with the text's description, our outputs closely match it while producing simulation-ready sewing patterns. 
\subsection{Image-Conditioned Generation Additional Visualization}

\begin{figure*}
    \centering
    \includegraphics[width=\textwidth]{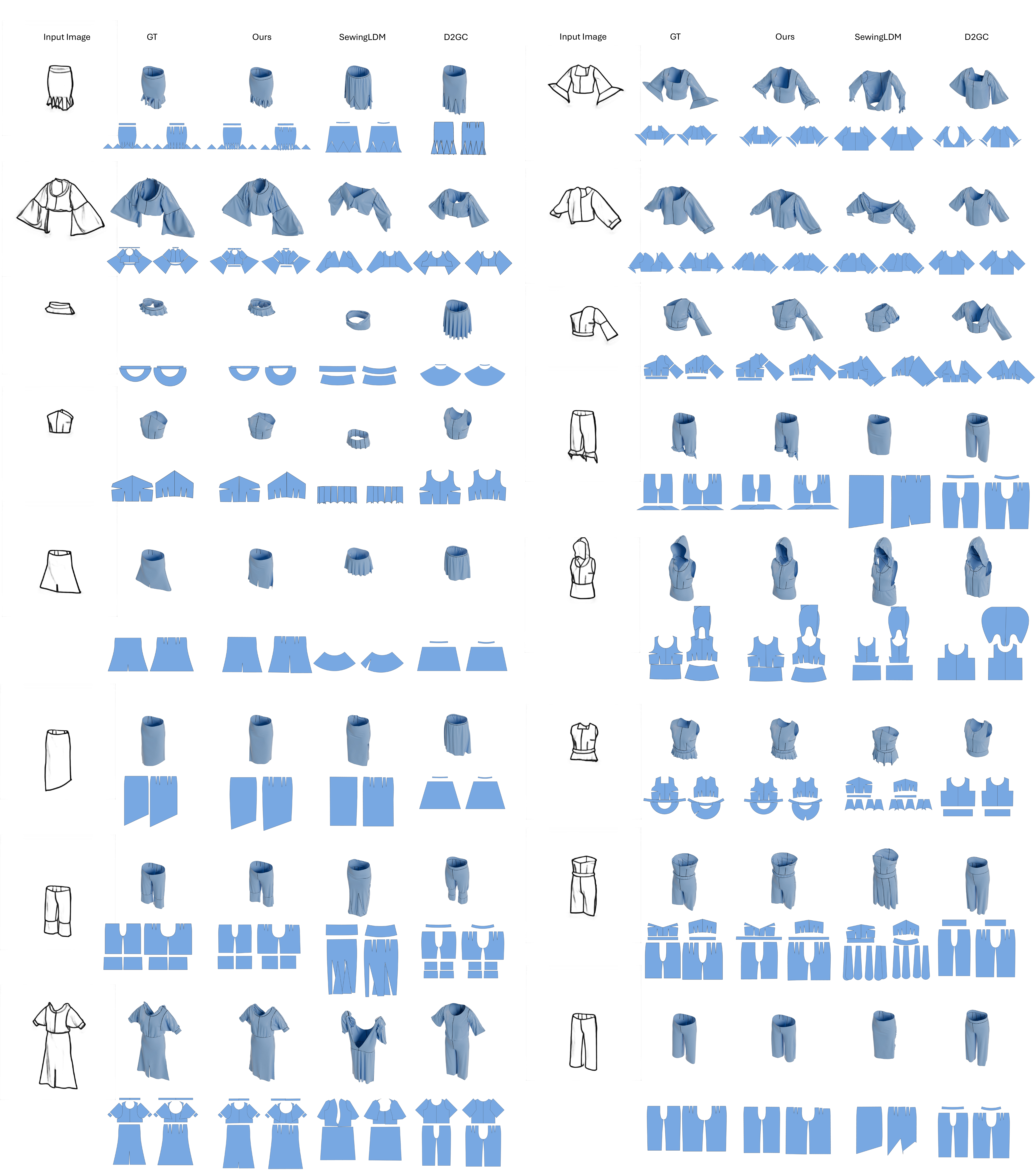}
    \caption{\textbf{Image-Conditioned Generation: Additional Visualization on Garment Sketches Dataset.}}
    \label{fig:linart_supp}
\end{figure*}

\begin{figure*}
    \centering
    \includegraphics[width=\textwidth]{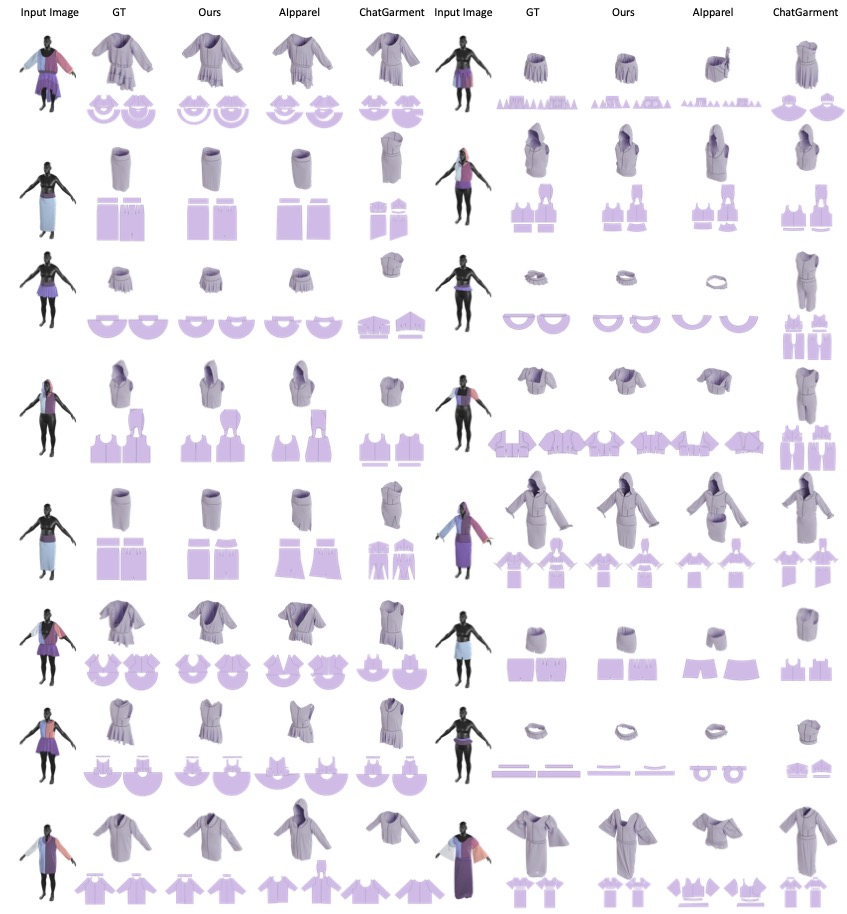}
    \caption{\textbf{Image-Conditioned Generation: Additional Visualization on GCDV2 Dataset.}}
    \label{fig:gcdv2_supp}
\end{figure*}

We showcase additional comparisons of our image-conditioned GPF model against baselines on the GCDv2 and Garment Sketches datasets. The results are shown in~\autoref{fig:gcdv2_supp} and~\autoref{fig:linart_supp}. On both datasets, our model consistently generates garments that are better aligned with the input sketch and the ground-truth garment. On the other hand, baseline generation can include patterns that do not drape correctly (Sketch: first and second rows of SewingLDM, GCDv2: first and last rows of AIpparel), result in incorrect garment styles (Sketch: third and seventh rows of Design2GarmentCode, GCDv2: last and second to last rows of ChatGarment), or have incorrect panel shapes (Sketch: fifth row of SewingLDM, GCDv2: last row of AIpparel).

\subsubsection{Extending GPF to more Modalities}
We experiment with extending GPF to multiview images. We use the front and back renderings from the \textit{GCDv2} and \textit{Garment Sketches} datasets as conditioning for the GPF. To pass into the model, we average the DINOV2 features of the front and back images for each token. \autoref{tab:multi_view_gen} shows the sewing pattern metric comparison against GPF with single-view input. Multiview conditioning consistently improves geometry reconstruction metrics, indicating that additional images provide useful constraints for generation.
% toward the ground truth, demonstrating the effectiveness of our fine-tuning scheme across different modalities. 

\subsubsection{Additional Garment Interpolation Results}
\begin{figure*}
    \centering
    \includegraphics[width=0.95\textwidth, trim={0 4.2cm 0 0},clip]{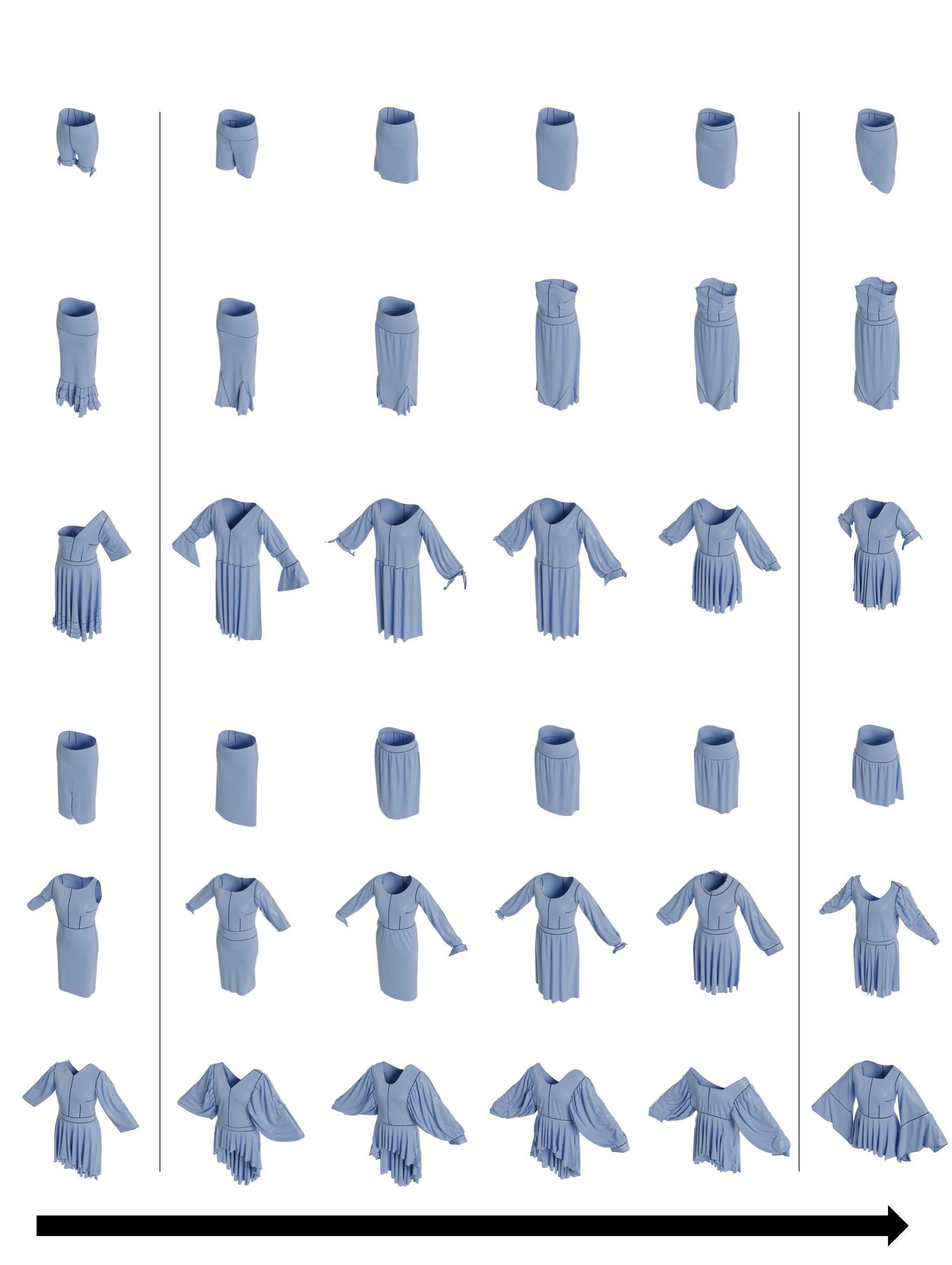}
    \vspace{-2em}
    \caption{\textbf{Garment Interpolation Results.}}
    \vspace{-5em}
    \label{fig:interpolation_supp}
\end{figure*}

We showcase additional results when interpolating between two generated garments from GPF in~\autoref{fig:interpolation_supp}. Our interpolation enables distinct garment style transitions (\eg pants-to-skirt: first row, single-sleeve-to-multi-sleeve (third, fifth rows)) and gradual size variations in different components of the garment (\eg larger sleeve (last row), shorter (fourth row), and longer skirts (second row)). These results demonstrate the representation effectiveness of our bidirectional garment particles representation.

\subsubsection{Additional Sewing Pattern Editing Results}
\begin{figure*}
    \centering
    \includegraphics[width=\textwidth]{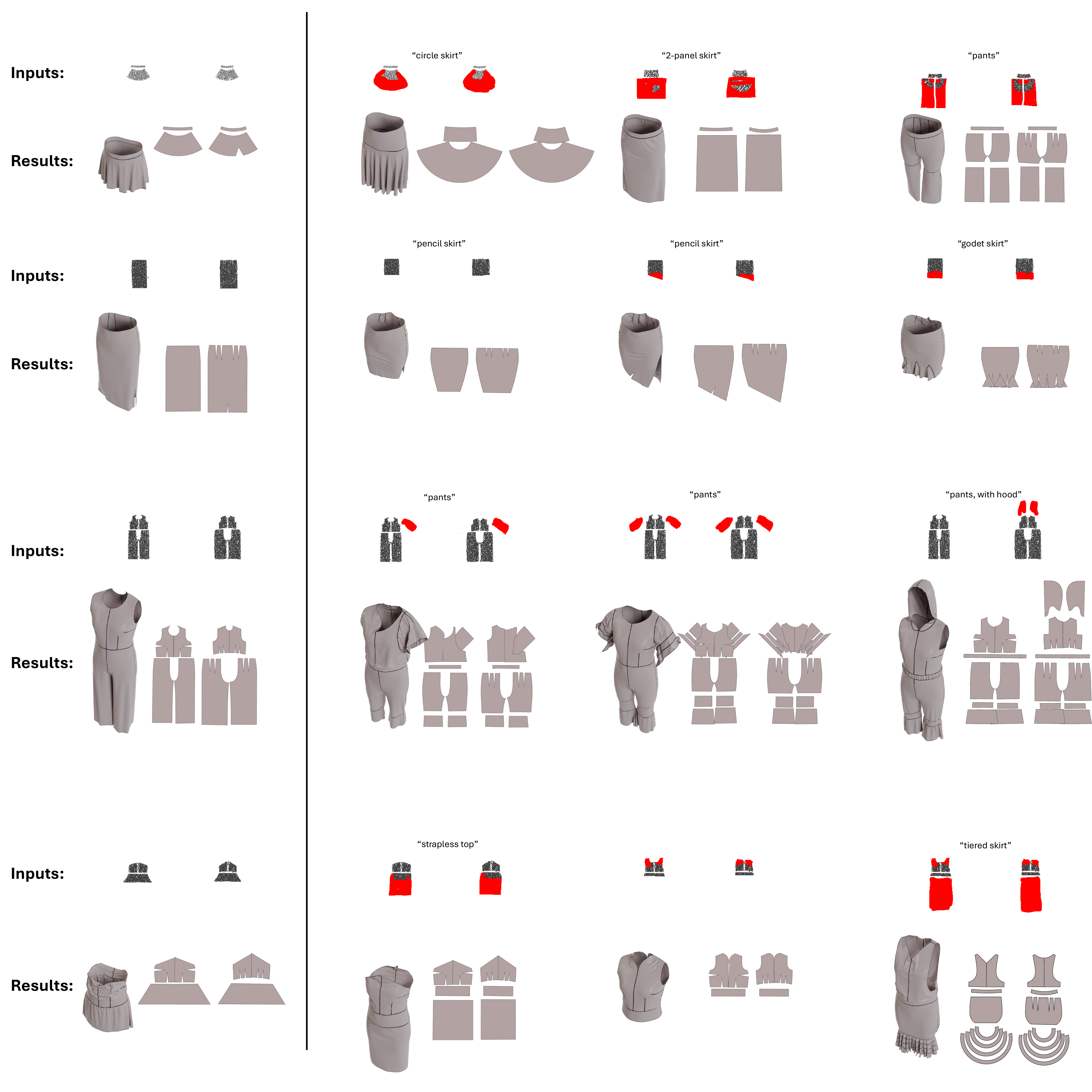}
    \caption{\textbf{Additional Sewing Pattern Editing Results.} Each row showcases a modified sewing pattern of the garment asset on the left. The red paint indicates users' input. The modified sewing pattern, combined with an optional text prompt, guides the generation of GPF garments. The generated garment asset after draping and its sewing pattern are shown below the inputs.}
    \label{fig:pattern_supp}
\end{figure*}

We showcase additional sewing pattern editing results in~\autoref{fig:pattern_supp}. The left shows the original garment generated using GPF. The left shows garments generated after editing the input sewing pattern, with red paint indicating users' input. The results show that our generated garments closely follow users' edits to the 2D sewing pattern, while filling in missing details when the input is coarse. We optionally use text as an additional control signal to guide the generation. However, because the guiding objective is agnostic of panel boundaries, we cannot control the number of panels generated. 
\subsubsection{Additional Silhouette-conditioned Garment Generation Results}
\begin{figure*}
    \centering
    \includegraphics[width=\textwidth]{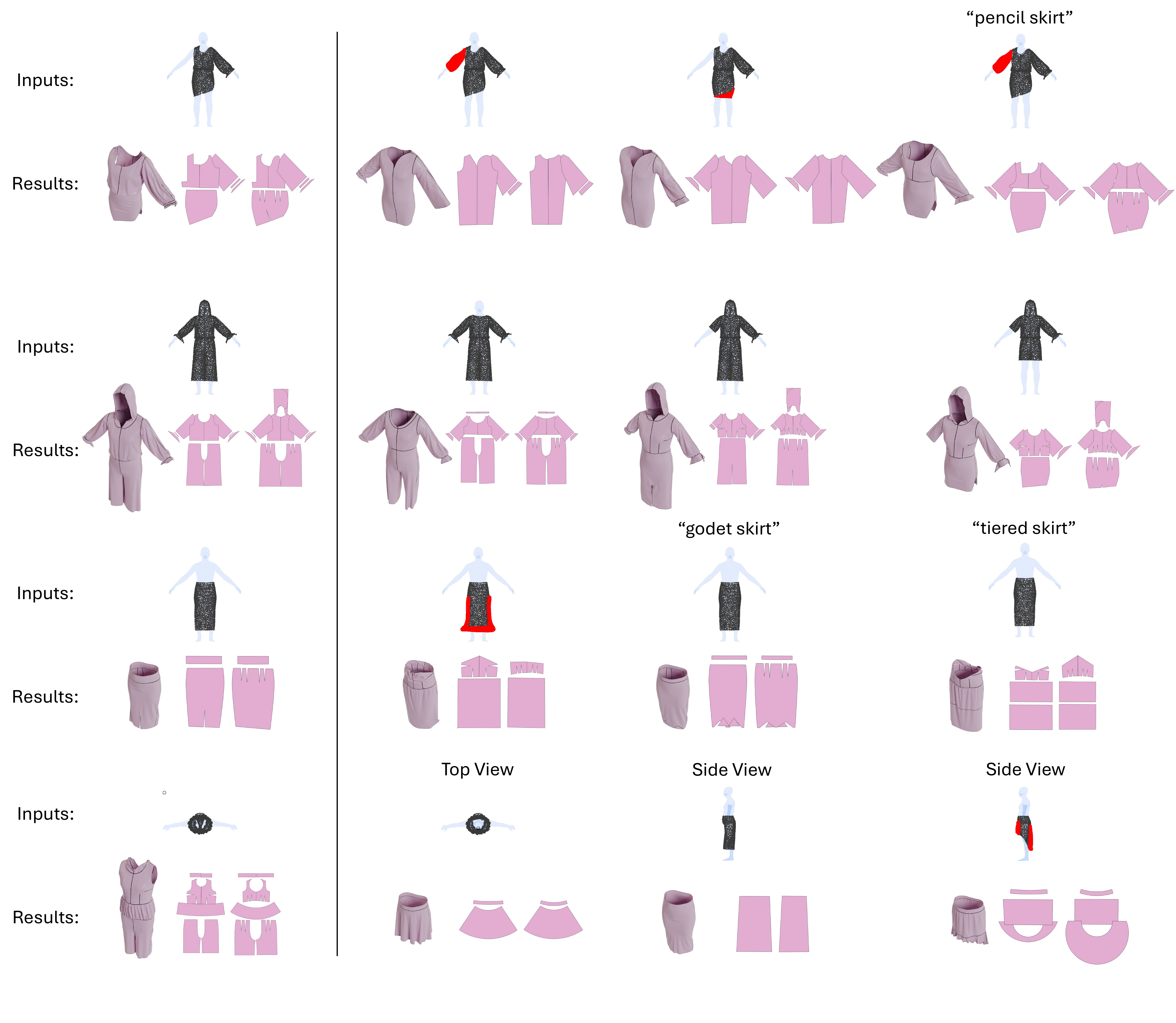}
    \caption{\textbf{Additional Silhouette Editing Results.} Each row showcases a garment asset generated by GPF conditioned on the silhouette shown in the input. An additional text prompt is also fed into GPF for an extra constraint. The leftmost column shows the initial garment generated from GPF, from which the silhouette edits are performed. The red paint indicates a newly added silhouette using our 2D interface. The last row shows silhouettes from different views that are used as conditioning. }
    \label{fig:silhouette_supp}
\end{figure*}

\autoref{fig:silhouette_supp} shows more results on silhouette-conditioned garment generation. The left column shows the original garment assets generated from GPF. We show three different edits on the silhouette, either erasing or adding, using our 2D user interface. The last row shows silhouette edits applied to different camera angles, including a top view where pants are converted to a skirt, and a side view where a skirt is widened and asymmetrized. We also optionally provide text as guidance to control the generated garment style. Our generated garments match the given silhouette while producing plausible draping results. 

\subsubsection{Additional Point-cloud-conditioned Garment Generation Results}
\begin{figure*}
    \centering
    \includegraphics[width=\textwidth]{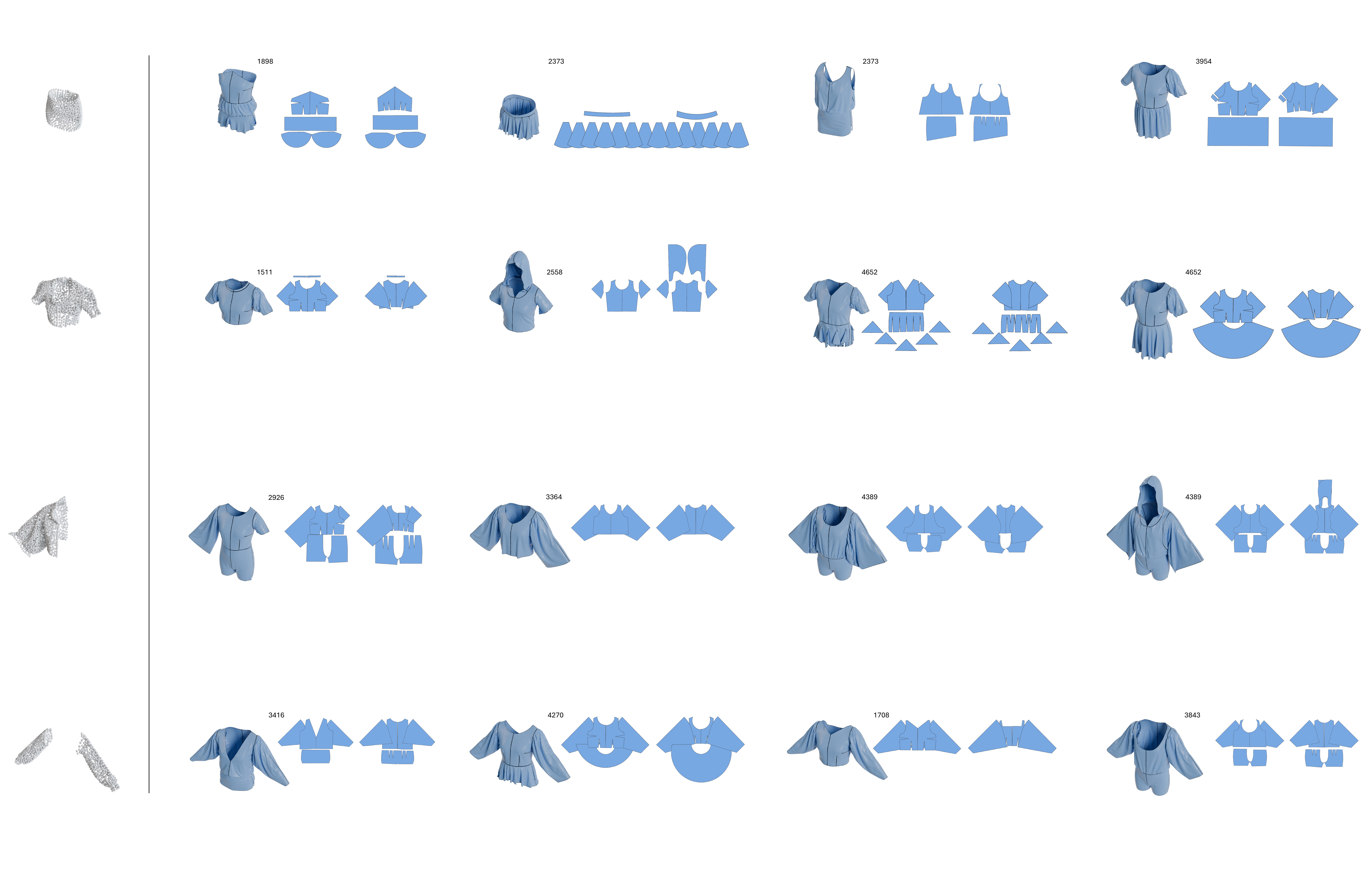}
    \caption{\textbf{Additional Point-cloud-conditioned Generation Results.}}
    \label{fig:3d_completion_supp}
\end{figure*}

\autoref{fig:3d_completion_supp} shows additional point-cloud-conditioned garment generation results. The models complete the partial 3D garment point cloud shown on the left. For each row we show three variations of possible completions. The number associated with each example shows the number of points we use for GPF generation. We see that as we increase the number of points, the complexity of the generated garment generally increases. 

\subsection{Human \& VLM Study}
\begin{table}[h]
\centering 
\begin{tabular}{ccccc}
\toprule
\text{AIpparel}&\text{SewingLDM}&\text{D2G}&\text{ChatGarment}&\text{Ours}\\\hline
923.8&985.1&1048.5&1060.4&\textbf{1065.4}\\\bottomrule
\end{tabular}
\caption{\textbf{Human Study ELO Ranking.} (Higher is better)}
\label{tab:human_study}
\end{table}
We conduct a human study on Amazon Mechanical Turk (440 responses) comparing the text alignment, aesthetic quality, and physical plausibility of text-based garment generation across baselines. Specifically, we present two garment renderings from separate models and ask users to select the garment with higher aesthetic appeal, physical plausibility, and text alignment. All the garments are generated with a selected set of 18 text prompts. After collecting all the responses, we compute ELO rankings following~\cite{wu2023gpteval3d}, and the results are shown in~\autoref{tab:human_study}. Our method achieves the highest score, indicating overall better alignment with text prompt, physical plausibility, and aesthetics compared with the baselines.

\begin{table*}[h]
\centering
\begin{tabular}{lccc}
\toprule
\textbf{Method} & \textbf{Garment Aesthetics} & \textbf{Text-Prompt Alignment} & \textbf{Physical Plausibility} \\
\midrule
AIpparel              & 821.6           & 888.2           & 857.1    \\
SewingLDM             & 993.3           & 1042.2          & 963.1           \\
D2GC    & 1060.9          & 1084.0          & 1030.5          \\
ChatGarment           & 1040.3          & 858.6           & 1048.3          \\
Ours                  & \textbf{1168.6} & \textbf{1143.8} & \textbf{1126.5} \\
\bottomrule
\end{tabular}
\caption{\textbf{VLM Evaluation Results}. (ELO rankings, higher is better)}
\label{tab:vlm_eval}
\end{table*}
For a more detailed analysis, we also conduct a VLM study, adapting the setup from GPTEval3D~\cite{wu2023gpteval3d} to evaluate the same three criteria separately. We use Gemini-2.5-flash as our VLM model. The ELO ranking after running the study is shown in~\autoref{tab:vlm_eval}. We see that the general ranking trend aligns with that from the human study, but reveals variations in the baselines when evaluating the three criteria separately. For example, while Chatgarment is second-best in physical plausibility, its text-prompt alignment is poor. In the meantime, Design2GarmentCode achieves a good balance among the three criteria.
% \begin{figure}[t]
%     \centering
%     \includegraphics[width=0.8\linewidth]{figures/fabrication_example.png}
%     \caption{\textbf{Fabricated Garments}}
%     \label{fig:fabrication}
% \end{figure}
\begin{figure*}[ht]
    \centering
    \includegraphics[width=\textwidth]{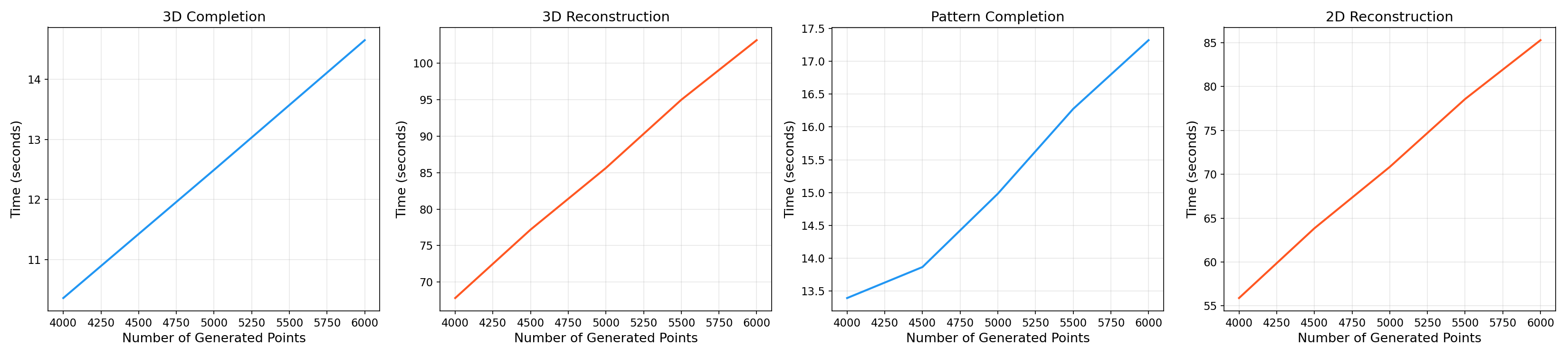}
    \caption{\textbf{Runtime Analysis for Garment Editing Tasks.}}
    \label{fig:runtime}
\end{figure*}
\subsection{Runtime Analysis}
\begin{table}[h]
\centering 

\begin{tabular}{cccc} 
\toprule
\text{AIpparel}&\text{D2GC}&\text{ChatGarment}&\text{Ours (GPF/PPF)}\\\hline4.59s&29.52s&14.24s&\textbf{4.01s}(2.48s/1.53s)\\
\bottomrule
\end{tabular}
\caption{\textbf{Garment Generation Runtime Comparison.} (Seconds, lower the better.)}
\label{tab:runtime}
\end{table}
\subsubsection{Generation Runtime Analysis}
In~\autoref{tab:runtime}, we compare with baselines the garment generation time(seconds) averaged over our test-set (1024 samples) using 100 denoising steps each. Our method achieves a similar runtime compared to AIpparel, and is much faster than ChatGarment and Design2GarmentCode since they require external LLM queries.

\subsubsection{Editing Runtime Analysis}
We also report garment editing runtime for the different tasks we showcased in the paper. Because the DPS algorithm's runtime depends on hyperparameters such as the input number of points, we plot the total runtime of DPS with different number of input points, given the same loss and observations. \autoref{fig:runtime} shows the runtime plot for the four garment editing tasks we shows in the paper. Specifically, 3D \& pattern reconstructions use EMD as their losses and completions use Chamfer Distance. In general, the runtime grows roughly linearly as the number of points grows. This suggests that the main time-bottleneck comes from CD or EMD computation instead of our network forward pass.

\subsection{Out-of-domain Evaluation}
\begin{table}[h]
\centering 
\begin{tabular}{ccccc}
\toprule
\text{DMap}&\text{GarmentRecovery}&\text{D2GC}&\text{ChatGarment}&\text{Ours}\\\hline
18.729&10.830&14.005&\textbf{7.104}&7.813\\\bottomrule
\end{tabular}
\caption{\textbf{Evalution on 4DDress Dataset} (Chamfer Distance, lower is better)}
\label{tab:4ddress}
\end{table}

To evaluate our method's ability to generalize to out-of-domain input data, we quantitatively evaluated our method against baselines on a subset of the 4DDress~\cite{wang20244ddress} dataset for the task of image-to-sewing-pattern reconstruction. \autoref{tab:4ddress} shows the comparison with both 3D garment reconstruction baselines, such as DMap~\cite{dmap_2025} and GarmentRecovery~\cite{li2024garment}, as well as sewing pattern generation baselines like ChatGarment~\cite{bian2024chatgarment} and Design2GarmentCode~\cite{zhou2024design2garmentcode}. Compared to the baselines, our method achieves comparable reconstruction accuracy compared to the state-of-the-art method while outperforming the optimization-based approaches, which heavily rely on human pose and input image segmentations.

\end{document}